%% file: konno2015_z2p2LAE_v8_revise_arXiv.tex
\documentclass[iop, apj, natbib]{emulateapj}
\usepackage{graphicx}
\usepackage{amsmath}
\usepackage{amssymb}
\usepackage{color}
\usepackage{multirow}
\usepackage{rotating}
\usepackage{booktabs}
\usepackage{times}
\usepackage{natbib}

\usepackage{comment}

\newcommand{\zi}{\textit{z}}
\newcommand{\ar}{\arcsec}
\newcommand{\lya}{Ly$\alpha$}
\newcommand{\HI}{\textsc{Hi}}

\slugcomment{Submission : December 7, 2015, Accepted : March 14, 2016}

\shorttitle{$\zi = 2$ \lya\ LF at the Bright and Faint Ends}
\shortauthors{Konno et al.}

\begin{document}

\title{Bright and Faint Ends of \lya\ Luminosity Functions at $\zi = 2$ Determined by the Subaru Survey:\\ 
Implications for AGN, Magnification Bias, and ISM \HI\ Evolution}
 
\author{Akira Konno\altaffilmark{1,2}, 
	Masami Ouchi\altaffilmark{1,3},
	Kimihiko Nakajima\altaffilmark{4},
	Florent Duval\altaffilmark{1},\\
	Haruka Kusakabe\altaffilmark{2},
	Yoshiaki Ono\altaffilmark{1}, and
	Kazuhiro Shimasaku\altaffilmark{2,5}
        }

\altaffiltext{1}{Institute for Cosmic Ray Research, The University of Tokyo, Kashiwa-no-ha, Kashiwa 277-8582, Japan}
\altaffiltext{2}{Department of Astronomy, Graduate School of Science, The University of Tokyo, Hongo, Bunkyo-ku, Tokyo, 113-0033, Japan}
\altaffiltext{3}{Kavli Institute for the Physics and Mathematics of the Universe (Kavli IPMU), WPI, The University of Tokyo, Kashiwa, Chiba 277-8583, Japan}
\altaffiltext{4}{Observatoire de Gen\`{e}ve, Universit\`{e} de Gen\`{e}ve, 51 Ch. des Maillettes, 1290 Versoix, Switzerland}
\altaffiltext{5}{Research Center for the Early Universe, Graduate School of Science, The University of Tokyo, Hongo, Bunkyo-ku, Tokyo, 113-0033, Japan}

\email{konno@icrr.u-tokyo.ac.jp}

\begin{abstract}
We present the \lya\ luminosity functions (LFs) derived by our deep Subaru narrowband survey
that identifies a total of 3,137 \lya\ emitters (LAEs) 
at $\zi = 2.2$ in five independent blank fields. 
The sample of these LAEs is the largest, to date, and covers a very wide \lya\ luminosity range of 
$\log L_{\mathrm{Ly}\alpha} = 41.7-44.4$ erg s$^{-1}$.
We determine the \lya\ LF at $\zi = 2.2$ with unprecedented accuracies,
and obtain the best-fit Schechter parameters of
$L^{*}_{\mathrm{Ly}\alpha} = 5.29^{+1.67}_{-1.13} \times 10^{42}$ erg s$^{-1}$,
$\phi^{*}_{\mathrm{Ly}\alpha} = 6.32^{+3.08}_{-2.31} \times 10^{-4}$ Mpc$^{-3}$, and
$\alpha = -1.75^{+0.10}_{-0.09}$
showing a steep faint-end slope.
We identify a significant hump at the LF bright end 
($\log L_{\mathrm{Ly}\alpha} > 43.4$ erg s$^{-1}$).
Because all of the LAEs in the bright-end hump have
(a) bright counterpart(s) either in the X-ray, UV, or radio data,
this bright-end hump is not made by gravitational lensing magnification bias but AGNs.
These AGNs allow us to derive the AGN UV LF at $\zi \sim 2$ down to the faint magnitude limit
of $M_{\rm UV}\simeq -22.5$, and to constrain the faint-end slope of AGN UV LF,
$\alpha_{\rm AGN}=-1.2 \pm 0.1$,
that is flatter than those at $\zi > 4$.
Based on 
the \lya\ and UV LFs from our and previous studies,
we find the increase of \lya\ escape fraction $f^{\mathrm{Ly}\alpha}_{\mathrm{esc}}$
from $\zi \sim 0$ to $6$ by two orders of magnitude.
This large $f^{\mathrm{Ly}\alpha}_{\mathrm{esc}}$ increase
can be explained neither by the evolution of stellar population nor outflow alone, 
but the evolution of neutral hydrogen \HI\ density in inter-stellar medium
that enhances dust attenuation for Ly$\alpha$ by resonance scattering.
Our uniform expanding shell models suggest that the typical \HI\ column density 
decreases from $N_{\textsc{Hi}} \sim 7 \times 10^{19}$ ($\zi \sim 0$) to $\sim 1 \times 10^{18}$ cm$^{-2}$ ($\zi \sim 6$)
to explain the large $f^{\mathrm{Ly}\alpha}_{\mathrm{esc}}$ increase.

\end{abstract}

\keywords{galaxies: evolution --- galaxies: formation --- galaxies: high-redshift --- \\
galaxies: luminosity function, mass function}

%%%%%%%%%%%%%%%%%%%%%%%%%%%%%%%%%%%%%%%%%%%%%%%%%%%%%%%%%%%%%%%%

\section{Introduction} \label{intro}

Deep 
narrowband and spectroscopic
observations identify \lya\ emitters (LAEs), most of which 
are continuum-faint star-forming galaxies with a prominent \lya\ emission line
(e.g., \citealt{Cowie98, Hu98, Rhoads00, Steidel00, Malhotra02, Hayashino04, Matsuda04, Taniguchi05, Iye06, Kashikawa06, Shimasaku06,
Dawson07, Gronwall07, Murayama07, Ouchi08, Finkelstein09, Guaita10, Adams11, Kashikawa12, Shibuya12, Yamada12, Konno14, Cassata15, Sobral15}).
LAEs are found at a wide redshift range of $\zi \sim 0 - 8$, and
\lya\ luminosity functions (LFs) of LAEs are used for probes of galaxy evolution and 
cosmic reionization (e.g., \citealt{Ouchi08, Ouchi10, Kashikawa11, Shibuya12, Konno14}).
In \lya\ LFs, there is an important characteristics at the faint end.
Galaxies at the faint end dominate in abundance, and 
faint-end slopes of \lya\ LFs are determined by mass, star-formation activities, physical conditions of inter-stellar medium (ISM), and feedback effects
that are key for understanding galaxy evolution (e.g., \citealt{Santos04, Rauch08}).
Although \lya\ LFs at various redshifts have been derived by previous observations,
faint-end slopes of the \lya\ LFs are poorly constrained
in contrast with those of UV LFs (e.g., \citealt{Reddy09, Oesch10, Hathi10, Sawicki12, Alavi14, Bouwens15, Parsa15}).
The faint-end LF slopes are quantified with $\alpha$, one of the three Schechter function parameters \citep{Schechter76},
depending on the rest of two parameters, characteristic \lya\ luminosity $L^{*}_{\mathrm{Ly}\alpha}$ and density $\phi^{*}_{\mathrm{Ly}\alpha}$.
Previous observational studies report $\alpha$ values for $\zi = 2-3$ Ly$\alpha$ LFs (e.g., \citealt{Cassata11}),
assuming a fixed parameter of $L^{*}_{\mathrm{Ly}\alpha}$ or $\phi^{*}_{\mathrm{Ly}\alpha}$.
There are some studies that constrain $\alpha$ values with no assumptions (e.g., \citealt{Gronwall07, Hayes10}),
but the uncertainties of the Schechter parameters are very large due to the small number of LAEs.
Although $\alpha$ is a parameter depending on $L^{*}_{\mathrm{Ly}\alpha}$ and $\phi^{*}_{\mathrm{Ly}\alpha}$,
so far, none of the observational studies have determined $\alpha$ simultaneously with $L^{*}_{\mathrm{Ly}\alpha}$ and $\phi^{*}_{\mathrm{Ly}\alpha}$
due to the small statistics of LAEs whose Ly$\alpha$ luminosity range is limited.
In theoretical studies, \cite{Gronke15} predict the three Schechter function parameters of \lya\ LFs
at $\zi = 3-6$, based on the measurements of UV LFs and \lya\ EW probability distribution functions (PDFs),
and argue that the faint-end slopes of the Ly$\alpha$ LFs are steeper than those of the UV LFs.

Another important characteristics of \lya\ LFs is found at the bright end.
The bright-end LFs are key for understanding 
massive-galaxy formation as well as faint active galactic nucleus (AGN; e.g., \citealt{Gawiser06, Ouchi08, Zheng13}).
Here, we define faint AGNs as AGNs whose LFs overlap with non-AGN galaxy LFs 
in the luminosity ranges.
The faint AGNs may play an important role in contributing to the UV radiation background (e.g., \citealt{Giallongo15}).
Faint AGNs are useful probes for quasar fueling lifetime, feedback, and duty cycle (e.g., \citealt{Hopkins06, Fiore12}).
Faint AGNs are spectroscopically identified
for most LAEs at $\zi \sim 3-4$ in the bright-end \lya\ LF at $\log L_{\mathrm{Ly}\alpha} \gtrsim 43.5$ erg s$^{-1}$ 
\citep{Gawiser06,Ouchi08}.
The bright-end LF includes an interesting physical effect, magnification bias.
The magnification bias effect boosts luminosities of high-$\zi$ galaxies by the gravitational lensing magnification
given by foreground massive galaxies, and flattens the bright-end LFs
(e.g. \citealt{Mason15}; see Figure 3 of \citealt{Wyithe11}).
In the observational studies, humps of the bright-end Ly$\alpha$ LF are found at
$\zi = 3-7$ \citep{Gawiser06,Ouchi08,Matthee15}.
In order to estimate the contributions of faint AGNs to the bright end LFs, 
it is important to investigate the properties of the bright-end galaxies 
with deep multiwavelength data such as X-ray, UV, and radio images. 

The intermediate redshift range of $\zi \sim 2-3$ is the best for investigating 
faint- and bright-end \lya\ LFs. This is because $\zi \sim 2-3$ is the lowest
redshift range where \lya\ emission fall in the optical observing window,
which allows us to identify very faint LAEs as well as
a large number of bright LAEs by fast optical surveys.
Moreover, because the number densities of AGNs peak at $\zi \sim 2-3$,
the effect of faint AGNs would clearly appear at the \lya\ LF bright end. 
By these reasons, in the past few years,
various surveys have been conducted to study LAEs at $\zi \sim 2-3$.
Although the \lya\ LFs at $\zi \sim 3$ are well determined (e.g. \citealt{Gronwall07,Ouchi08}),
those at $\zi \sim 2$ are derived with uncertainties larger than those at $\zi \sim 3$ due to
difficulties of $\mathcal{U}$-band observations at $\sim 3000-4000$\AA\ to which \lya\ emission lines of $\zi \sim 2$ objects are redshifted.
Thus, the evolution of \lya\ LFs from $\zi \sim 2$ to $3$ is under debate.
\cite{Nilsson09} first claim that there is a possible evolution of 
LAE number densities
between $\zi = 2.25$ and $\sim 3$
albeit with the large uncertainties originated from the small sample.
Subsequent studies have identified $\zi \sim 2$ LAEs by narrowband imaging and spectroscopic observations,
and discussed the evolution of \lya\ LFs and the integrations of \lya\ LFs, luminosity densities (LDs), at $\zi = 2-3$.
\cite{Cassata11} and \cite{Blanc11} have carried out blank-field spectroscopy for LAEs at $2 < \zi < 6.6$ and $1.9 < \zi < 3.8$, respectively,
and concluded no evolution of the \lya\ LDs from $\zi = 2$ to  $3$. 
On the other hand, \cite{Ciardullo12} show that the \lya\ LF evolves from $\zi = 2.1$ to $3.1$ significantly
by the narrowband imaging surveys in ECDF-S (see also \citealt{Guaita10}).
Because the $\zi \sim 2$ LAE samples of these studies are limited in the LAE numbers
(that are equal to or less than several hundreds)
and the \lya\ luminosity dynamic range (that is a factor of $\sim10$),
these discrepancies may be raised by the sample variances and the differences of
\lya\ luminosity coverages.

Evolution of \lya\ LFs at
$\zi \lesssim 2-3$ is 
also discussed extensively.
\cite{Deharveng08} claim that there is a substantial drop in the \lya\ LFs from $\zi \sim 3$ to $\sim 0.3$
(see also \citealt{Cowie10, Cowie11, Barger12, Wold14}).
A similar evolutionary trend can be found in the \lya\ escape fraction at $\zi \sim 0 - 6$ (e.g., \citealt{Hayes11}, \citealt{Blanc11}, \citealt{Zheng13})
that is defined by the ratio of the observed to the intrinsic \lya\ fluxes.
The physical origin of the rapid evolution may be dust attenuation within galaxies.
From the observations of UV-continuum slope of Lyman break galaxies (LBGs),
dust extinction, $E(B-V)$, 
decreases 
toward higher redshift (e.g., \citealt{Bouwens15}).
Because the \lya\ escape fraction clearly depends on $E(B-V)$ (e.g., \citealt{Kornei10, Atek14}),
dust extinction would explain the rapid evolution of the \lya\ LF and \lya\ escape fraction.
To understand the major physical mechanisms related to the \lya\ escape processes at high-$\zi$
and its dependence on redshift, determining \lya\ LFs at $\zi \sim 2$ is important.

In this paper, we present our analyses and results of the \lya\ LFs at $\zi = 2.2$
based on our large LAE sample given by Subaru narrowband observations (\citealt{Nakajima12, Nakajima13}; see also \citealt{Kusakabe15}).
This sample contains 3,137 LAEs 
at $\zi =2.2$ with a wide \lya\ luminosity range of 
$41.7 \leqslant \log L_{\mathrm{Ly}\alpha} \leqslant 44.4$ erg s$^{-1}$,
and enables us to examine the faint+bright ends and the evolution of \lya\ LFs.
We describe the details of our observations and our $\zi = 2.2$ LAE candidate selection in Section \ref{sec:observation}.
We derive the \lya\ LFs at $\zi = 2.2$, and compare the LFs with those of previous studies in Section \ref{sec:LF}.
We investigate the \lya\ LF and LD evolution from $\zi \sim 2$ to $3$, and 
extend the discussion to the wider redshift range of $\zi \sim 0-8$ in Section \ref{sec:evoLya}.
We finally discuss the physical origins of the
bright-end of our $\zi = 2.2$ \lya\ LFs,
and the \lya\ LD evolution at $\zi \sim 0-8$ in Section \ref{sec:discuss}.
Throughout this paper, we adopt AB magnitudes \citep{Oke74}
and concordance cosmology with a parameter set of 
($h$, $\Omega_\mathrm{m}$, $\Omega_\Lambda$, $\sigma_8$) = (0.7, 0.3, 0.7, 0.8) consistent 
with the nine-year \textit{WMAP} and \textit{Planck} 2015 results \citep{Hinshaw13, Planck15}.

%%%%%%%%%%%%%%%%%%%%%%%%%%%%%%%%%%%%%%%%%%%%%%%%%%%%%%%%%%%%%%%%
\section{Observations and Sample Selection} \label{sec:observation}

\subsection{\textit{NB387} Observations} \label{sec:image}

We have conducted a deep and large-area narrowband imaging survey for $\zi = 2.2$ LAEs with Subaru/Suprime-Cam \citep{Miyazaki02}.
For these observations, we have developed a new narrowband filter, \textit{NB387},
with a central wavelength, $\lambda_\mathrm{c}$, of $3870 \mathrm{\AA}$ and
an FWHM of $94 \mathrm{\AA}$ to identify LAEs in the redshift range of $\zi = 2.14 - 2.22$.
With our \textit{NB387} filter, we have observed five independent blank fields, the Subaru/\textit{XMM-Newton} Deep Survey (SXDS) field
\citep{Furusawa08}, the Cosmic Evolution Survey (COSMOS) field \citep{Scoville07}, the \textit{Chandra} Deep Field South (CDFS; \citealt{Giacconi01}), 
the \textit{Hubble} Deep Field North (HDFN; \citealt{Capak04}), and the SSA22 field (e.g., \citealt{Steidel00}),
in 2009 July 20 and December $14-16$, $19-20$.
The SXDS field consists of five subfields of $\sim 0.2$ deg$^2$, SXDS-C, -N, -S, -E, and -W \citep{Furusawa08}.
We cover 
these five SXDS subfields, COSMOS, CDFS, HDFN, and SSA22 by one pointing of Suprime-Cam whose field of view is $\sim0.2$ deg$^2$.
We thus have \textit{NB387} imaging data in a total of nine pointing positions of Suprime-Cam. 
We summarize the details of our observations as well as image qualities in Table \ref{table:imaging_data}.
In this study, we do not use the data of SXDS-E subfield due to the poor seeing size of $\simeq 2\ar$ 
in FWHM of point-spread function (PSF; see Table\ref{table:imaging_data}).
During our observations, we have taken spectrophotometric standard stars
Feige34, LDS749B, and G93-48 \citep{Oke90} for photometric calibration.
Each standard star has been observed more than twice 
under the photometric condition with air masses of 1.1$-$1.3. 

\input{imaging_data}

\subsection{Data Reduction} \label{sec:reduc}

Our \textit{NB387} data are reduced with the Suprime-Cam Deep Field REDuction (SDFRED) package \citep{Yagi02, Ouchi04}.
The data reduction process includes
the subtraction of bias estimated with overscan regions,
flat fielding, distortion+atmospheric-dispersion correction,
cosmic-ray rejection, sky subtraction, image shifting, and stacking.
In cosmic-ray rejection process, we use \textit{LA.COSMIC} \citep{vanDokkum01}.
Before the image shifting, we mask out bad pixels and satellite trails.

After the stacking process, we calculate photometric zero points of the \textit{NB387} images 
from the standard-star data (see Section \ref{sec:image}).
We estimate the errors in the photometric zero points based on 
colors of stellar objects in the two-color diagram of 
$NB387$ and two adjacent broadbands in the blue and red sides of $NB387$
 (e.g., \textit{u$^*$}$-$\textit{NB387} and \textit{B}$-$\textit{NB387} in SXDS).
We compare the colors of stellar objects and the template 175 Galactic stars \citep{Gunn83},
and regard the offsets as the uncertainties.
The inferred uncertainties are $\lesssim 0.05$ mag, which are negligibly small 
for our study.

All of the \textit{NB387} images, except the SXDS-E data, have the PSF FWHM of $0\farcs7-1\farcs2$,
and reach the $5\sigma$ limiting magnitudes of $24.9-26.5$ in a $2\farcs0$-diameter circular aperture.
We summarize the qualities of these reduced \textit{NB387} images in Table \ref{table:imaging_data}.
We mask out the imaging regions that are 
contaminated with halos of bright stars, CCD blooming, 
and the low signal-to-noise ratio pixels near the edge of the images.
After the masking, the total survey area is 5,138 arcmin$^2$, i.e. $\simeq 1.43$ deg$^2$.
If we assume a simple top-hat selection function for LAEs whose redshift distribution is defined by the FWHM of \textit{NB387},
this total survey area corresponds to the comoving volume of $\simeq 1.32 \times 10^6$ Mpc$^3$.

In our analysis and LAE selection, we use archival
$\mathcal{U}$- and \textit{B}-band data 
as well as our \textit{NB387} images.
In the SXDS field, the \textit{u$^*$}- and \textit{B}-band data are taken with CFHT/MegaCam (S. Foucaud et al., in preparation)
and Subaru/Suprime-Cam \citep{Furusawa08}, respectively.
The \textit{u$^*$}- and \textit{B}-band images in the COSMOS field are obtained with CFHT/MegaCam \citep{McCracken10}
and Subaru/Suprime-Cam \citep{Capak07}, respectively.
We use VLT/VIMOS \textit{U}-band \citep{Nonino09}
and MPG 2.2m Telescope/WFI \textit{B}-band \citep{Hildebrandt06} images in CDFS 
(see \citealt{Kusakabe15} for more details),
and KPNO 4m Telescope/MOSAIC prime focus camera \textit{U}-band 
and Subaru/Suprime-Cam \textit{B}-band images in HDFN \citep{Capak04}.
In SSA22 field, we use the \textit{u$^*$}-band data of CFHT/MegaCam 
and \textit{B}-band data of Subaru/Suprime-Cam \citep{Hayashino04}.
The properties of these optical broadband data are also summarized in Table \ref{table:imaging_data}.
Note that in CDFS, \cite{Nakajima13} do not use the VLT/VIMOS \textit{U}-band image,
but only the MPG 2.2m Telescope/WFI \textit{U}-band image \citep{Gawiser06, Cardamone10}
that is significantly shallower than the VLT/VIMOS \textit{U}-band data.
The deep VLT/VIMOS \textit{U}-band image allows us to remove foreground contamination
efficiently, although the area coverage of VLT/VIMOS \textit{U}-band data is smaller than 
that
of MPG 2.2m Telescope/WFI \textit{U}-band data. We thus use the deep VLT/VIMOS \textit{U}-band 
image.

To measure colors of objects precisely, we align our \textit{NB387} images with the broadband data
using bright stellar objects commonly detected in the \textit{NB387} and the broadband images.
After the image alignment process, we match the PSF sizes of broadband and narrowband images
in each field,
referring to these stellar objects.

\subsection{Photometric Sample of $\zi =2.2$ LAEs} \label{sec:sample}

Our source detection and photometry are performed with SExtractor \citep{Bertin96}.
We use the PSF-homogenized images (Section \ref{sec:reduc}) to measure colors of objects.
We identify sources that are made of contiguous $>5$ pixels whose counts are
above the $>2\sigma$ brightness of the background fluctuations in our \textit{NB387} images.
We obtain a circular aperture magnitude of SExtractor's \texttt{MAG\_APER} 
with an aperture's diameter of $2\farcs5$ in the SXDS-W field,
$3\farcs0$ in the HDFN field, and $2\farcs0$ in the other fields,
and define a $5\sigma$-detection limit magnitude with the aperture size in each field.
The different aperture diameters are applied, because
the PSF sizes of the homogenized images in the SXDS-W and HDFN are large,
$1\farcs23$ and $1\farcs29$, respectively.
We use the aperture magnitudes to calculate colors of the sources, and adopt \texttt{MAG\_AUTO} of SExtractor for our total magnitudes.
All magnitudes of the sources are corrected for Galactic extinction of $E(B-V) = 0.020$, $0.018$, $0.008$, $0.012,$ and $0.08$
in the SXDS, COSMOS, CDFS, HDFN, and SSA22 fields, respectively \citep{Schlegel98}.
We thus obtain source catalogs that contain 42,995 (SXDS), 31,401 (COSMOS), 24,451 (CDFS), 36,236 (HDFN), and 8,942 (SSA22) objects
with the aperture magnitudes brighter than the $5\sigma$-detection limit magnitudes.

\begin{figure*}
\centering
\includegraphics[width=0.90\textwidth]{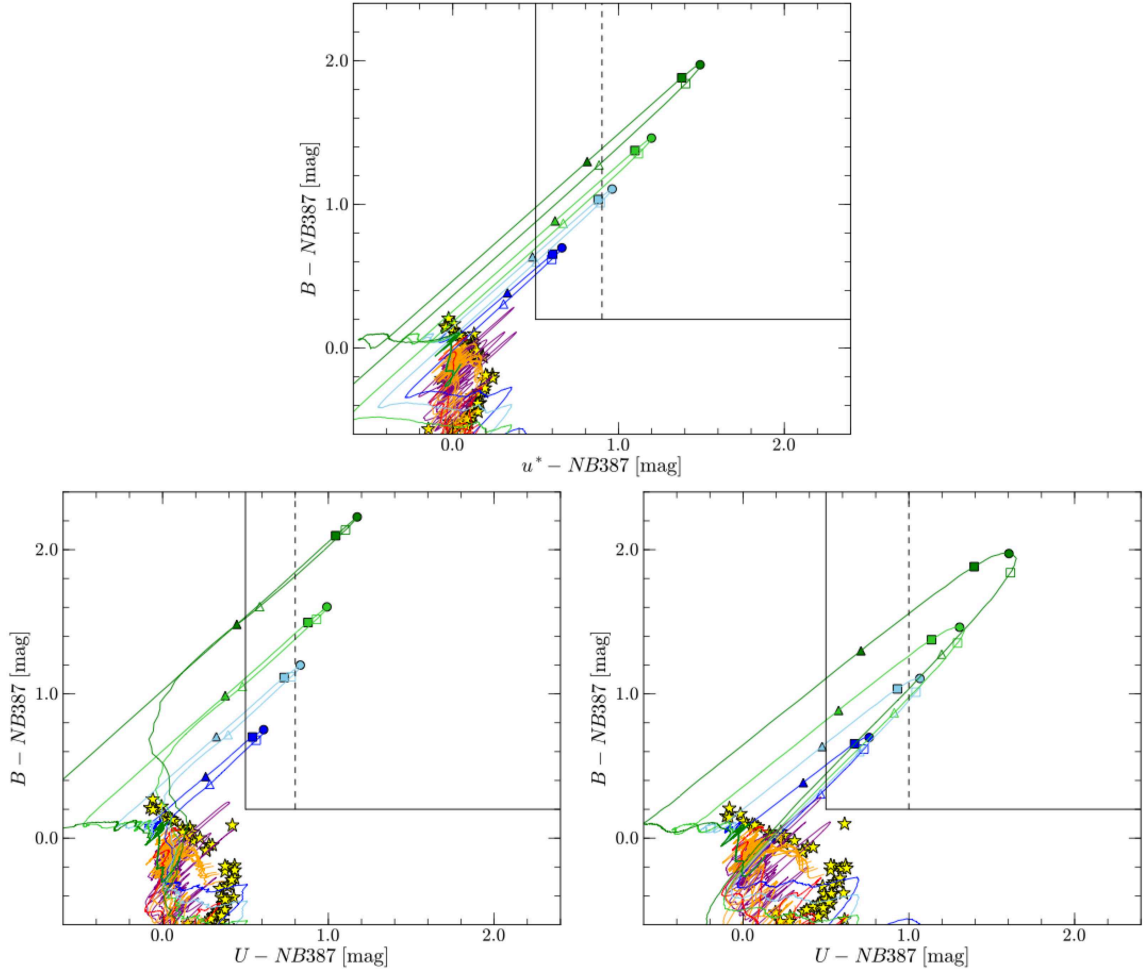}
\caption{
Two color diagrams 
for the selection of $\zi = 2.2$ LAEs:
\textit{B}$-$\textit{NB387} vs. \textit{u$^*$}$-$\textit{NB387} for SXDS, COSMOS, and SSA22 
(top);
\textit{B}$-$\textit{NB387} vs. \textit{U}$-$\textit{NB387} in CDFS 
(bottom left);
\textit{B}$-$\textit{NB387} vs. \textit{U}$-$\textit{NB387} in HDFN 
(bottom right).
The solid lines in blue, light blue, light green, and green represent 
the color tracks of redshifted model LAE SEDs with Ly$\alpha$ EW$_0$ $= 30, 60, 100,$ and $200\mathrm{\AA}$, respectively. 
These models are produced by using the \cite{Bruzual03} population synthesis model where we adopt a $30$ Myr simple stellar population with Salpeter IMF and adding a Ly$\alpha$ emission. 
We apply the \cite{Madau95} prescription to take into account the inter-galactic medium (IGM) absorption.
The symbols on these tracks correspond to $\zi = 2.14$ (filled triangles), $2.16$ (filled squares),
$2.18$ (filled circles), $2.20$ (open squares), and $2.22$ (open triangles).
The red and orange curves show the tracks of three elliptical (age of 2, 5, 13 Gyr) and six spiral (S0, Sa, Sb, Sc, Sd, and Sdm) galaxies
from the SWIRE template library \citep{Polletta07}, respectively.
The purple solid lines indicate six templates of nearby starburst galaxies \citep{Kinney96}.
These elliptical, spiral and starburst template galaxies are redshifted from $\zi = 0.0$ up to $2.0$ 
with a step of $\Delta \zi = 0.001$.
The yellow star marks are 175 Galactic stars given by \cite{Gunn83}.
The black solid and dashed lines represent the color criteria to select our $\zi = 2.2$ LAE candidates
whose \lya\ EWs are larger than $20-30\mathrm{\AA}$ and $60\mathrm{\AA}$, respectively.
}
\label{fig:2CDmodel_panel}
\end{figure*}

We select $\zi=2.2$ LAE candidates based on narrowband excess colors of 
$\mathcal{U} -$\textit{NB387} and \textit{B}$-$\textit{NB387},
in the same manner as \cite{Nakajima12} who present the first results of the \textit{NB387} observations in the SXDS field.
Here, $\mathcal{U}$ indicates \textit{u$^*$} or \textit{U}. 
Figure \ref{fig:2CDmodel_panel} presents two color diagrams of 
\textit{B}$-$\textit{NB387} versus
$\mathcal{U} -$\textit{NB387}.
In this figure, we plot colors of model galaxies and Galactic stars to define the selection criteria for $\zi = 2.2$ LAE candidates.
Based on Figure \ref{fig:2CDmodel_panel}, we apply the color criteria \citep{Nakajima12, Nakajima13, Kusakabe15} 
\begin{equation}
	\mathcal{U} - \textit{NB387} > 0.5 \ \mathrm{and} \ \textit{B} - \textit{NB387} > 0.2 	\label{eq:full}
\end{equation}
to obtain $\zi = 2.2$ LAE candidates whose 
rest-frame \lya\ equivalent width, EW$_0$, are EW$_0$ $\gtrsim 20-30 \mathrm{\AA}$.
After the visual inspection to remove spurious sources, such as 
ghosts, bad pixels, surviving cosmic rays (see \citealt{Nakajima12} for more details),
we identify 3,137 LAE candidates 
in our survey fields.
The sample of these LAE candidates is referred to as the full sample.
This is so far the largest LAE sample in the large area field surveys
(cf. 187 and 250 LAEs at $\zi \simeq 2.2$ with EW$_0 > 20 \mathrm{\AA}$ observed 
by \citealt{Nilsson09} and \citealt{Guaita10}, respectively).
We summarize the details of the full sample 
in Table \ref{table:sample}.

We make a subsample with the uniform criterion of  \lya\ EW$_0>60\mathrm{\AA}$
to compare the \lya\ LF at $\zi = 3.1$ of \cite{Ouchi08} (see Section \ref{sec:evoLF}),
and refer to the subsample as the EWgt60 sample.
We apply the color criteria of
\begin{align}
	\textit{u}^* - \textit{NB387} > 0.9 \ & \mathrm{and} \ \textit{B} - \textit{NB387} > 0.2  	\notag \\
								& \mathrm{in} \ \mathrm{SXDS,} \ \mathrm{COSMOS,} \ \mathrm{and} \ \mathrm{SSA22,}	\label{eq:601}	 	\\
	\textit{U} - \textit{NB387} > 0.8 \ &\mathrm{and} \ \textit{B} - \textit{NB387} > 0.2 		\notag	\\
							&  \mathrm{in} \ \mathrm{CDFS,}	\label{eq:602}	\\
	\textit{U} - \textit{NB387} > 1.0 \ &\mathrm{and} \ \textit{B} - \textit{NB387} > 0.2 		\notag	\\
							& \mathrm{in} \ \mathrm{HDFN}		\label{eq:603}
\end{align}
for the EWgt60 sample.
After the visual inspection, we obtain 985 LAE 
candidates for the EWgt60 sample
that is summarized in
Table \ref{table:sample}.

\input{sample}

%%%%%%%%%%%%%%%%%%%%%%%%%%%%%%%%%%%%%%%%%%%%%%%%%%%%%%%%%%%%%%%%
\section{Luminosity Function} \label{sec:LF}

\subsection{Contamination} \label{sec:cont}

We investigate the contamination sources of our LAE samples
that are low-$\zi$ emitters whose emission lines are redshifted to the bandpass of \textit{NB387}. 
The major strong emission that 
enters
into the \textit{NB387} bandpass is [\textsc{Oii}]$\lambda 3727$.
However, our survey area of 5,138 arcmin$^2$ (Section \ref{sec:reduc}) corresponds 
to the comoving volume of $1.22 \times 10^3$ Mpc$^3$
for [\textsc{Oii}] emitters at $\zi = 0.04$, which is three orders of magnitude smaller than the survey volume 
of our $\zi =2.2$ LAEs ($1.32 \times 10^6$ Mpc$^3$).
Moreover, the color criterion defined by Equation (\ref{eq:full}) corresponds to 
a relatively large rest-frame EW limit of $\gtrsim 70${\AA} for $z = 0.04$ [\textsc{Oii}] emitters.  
\cite{Ciardullo13} examine [\textsc{Oii}] LFs and EW distributions 
at $\zi \sim 0.1$ 
and find that the [\textsc{Oii}] EW distribution 
has an exponential scale of $8.0$\AA, 
which is significantly smaller than our selection criterion for [\textsc{Oii}] emitters (i.e., EW$_0$ $\sim 70$\AA).
Based on our survey parameters (see Sections \ref{sec:reduc}-\ref{sec:sample})
and the \citeauthor{Ciardullo13}'s [\textsc{Oii}] LF and EW distribution, 
the expected number of [\textsc{Oii}] emitters at $\zi = 0.04$ in our full sample is 
$\sim 3 \times 10^{-2}$.
Therefore, the probability of the [\textsc{Oii}] emitter contamination would be very small.
We further discuss the possibility that our bright sources would include
\textsc{Civ}$\lambda 1548$ and \textsc{Ciii}]$\lambda 1909$ emitters at $\zi \sim 1.5$.
These \textsc{Civ} and \textsc{Ciii}] emitters should be mostly AGNs,
because these emitters have to have a \textsc{Civ} or \textsc{Ciii}] EW
greater than $30$\AA\ to pass our selection criterion.
This EW value is significantly larger than the one of the star-forming galaxies.
Because in Section \ref{sec:AGN_LF}, we find that our AGN UV LF is consistent with
the previous SDSS measurements, only a negligibly small fraction of the $\zi \sim 1.5$ AGNs include our sample.

Nevertheless, spectroscopic follow-up observations for our LAEs have been conducted 
with Magellan/IMACS, MagE, and Keck/LRIS by \cite{Nakajima12}, \cite{Hashimoto13}, 
\cite{Shibuya14} and M. Rauch et al., in preparation.
A total of 43 LAEs are spectroscopically confirmed.
These spectroscopic observations find no foreground interlopers 
such as {\sc [Oii]} emitters at $z=0.04$ that show 
{\sc [Oiii]}5007 emission at $5200$\AA\ (see e.g. \citealt{Nakajima12}).
We note that these spectroscopic redshift confirmations are limited
to the bright LAEs with $\textit{NB387} \lesssim 24.5$,
and that the number of the faint LAEs confirmed by spectroscopy is small.
However, the contamination rate at the faint end is probably not high.
This is because the EW criterion of our selection corresponds to
$\sim 70$\AA\ for the major foreground faint emitters of $\zi = 0.04$ [\textsc{Oii}] emitters.
Most of these potential contamination sources do not pass this large EW limit, as discussed above.
Thus, the effects of contamination sources are negligibly small
in our LAE samples.

\subsection{Detection Completeness} \label{sec:detcomp}

\begin{figure}
\centering
\includegraphics[width=8.4cm]{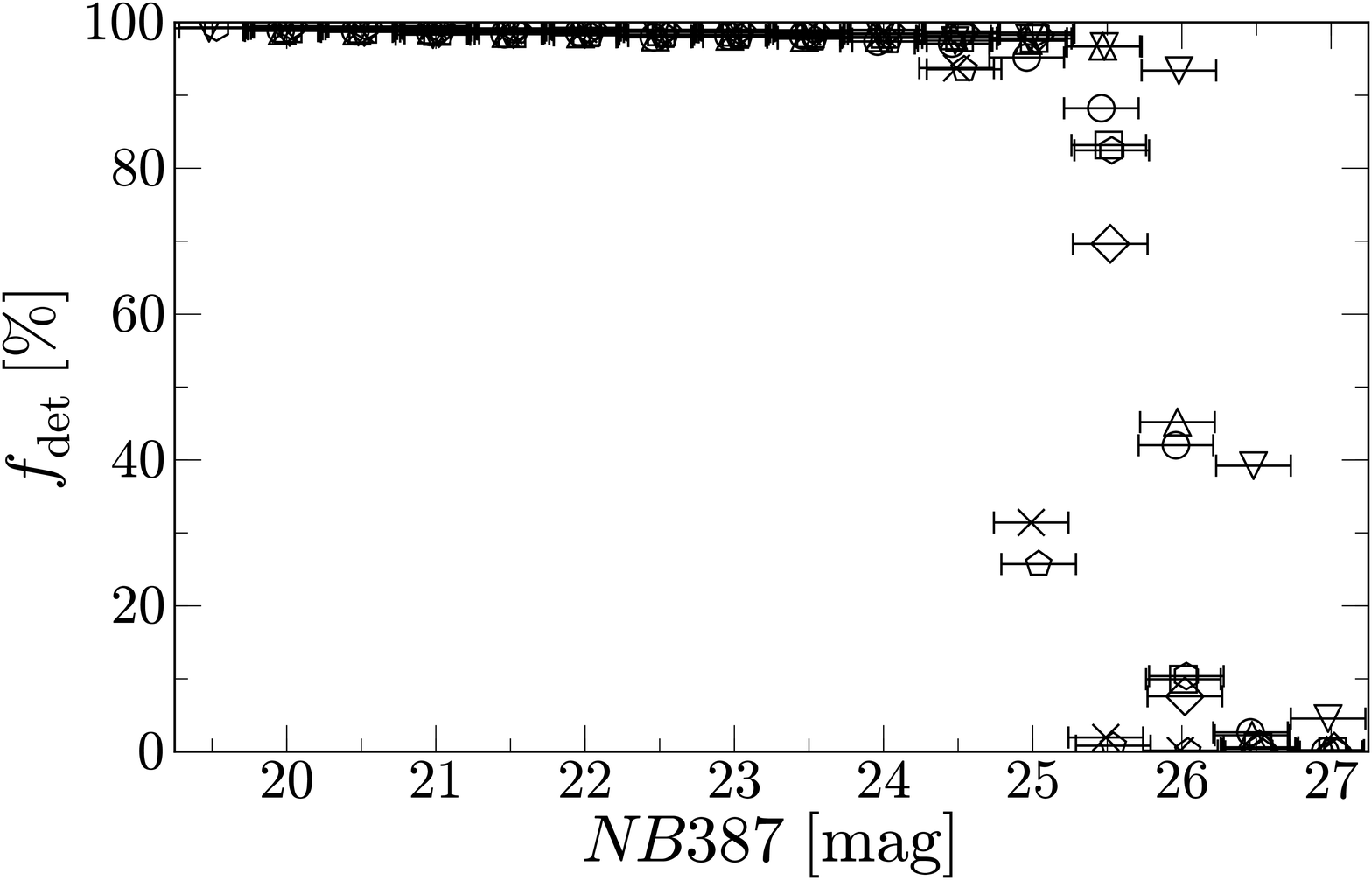}
\caption{Detection completeness, $f_\mathrm{det}$, of our \textit{NB387} images.
The symbols represent the completeness 
in a magnitude bin of $\Delta m = 0.5$ mag
for the SXDS-C (squares), SXDS-N (diamonds), SXDS-S (hexagons), SXDS-W (pentagons),
COSMOS (circles), CDFS (inverted triangles), HDFN (triangles), and SSA22 (cross marks) fields.
For presentation purposes, we slightly shift all the points along the abscissa.
} 
\label{fig:det_comp}
\end{figure}

We evaluate detection completeness in each field 
by Monte-Carlo simulations, following the procedures of \cite{Konno14}.
We randomly distribute a total of $\sim$5,000 artificial sources mimicking LAEs in each \textit{NB387} image,
and detect the artificial sources in the same manner as the real source identifications (Section \ref{sec:sample}).
Here, we assume that LAEs at $\zi = 2.2$ are point sources, and use profiles obtained by
the stack of 500 bright point sources in each \textit{NB387} image.
We define the detection completeness as a fraction of the numbers of the detected artificial sources 
to all of the input artificial sources.
We obtain the detection completeness as a function of \textit{NB387} magnitude,
repeating this process with various magnitudes of the input artificial sources.
Figure \ref{fig:det_comp} shows the results of these Monte-Carlo simulations.
We find that the detection completeness is typically $\gtrsim 90$\% for relatively bright sources (\textit{NB387} $< 24.5$) in all fields,
and $\sim 50$\% at around the $5\sigma$ limiting magnitude of \textit{NB387} in each field (see Table \ref{table:imaging_data}).

\subsection{Cosmic Variance} \label{sec:cosvar}

To include field-to-field variation in the error bar of our \lya\ LFs, we calculate 
the cosmic variance uncertainty, $\sigma_\mathrm{g}$, with
\begin{equation}
\sigma_\mathrm{g} = b_\mathrm{g} \sigma_\mathrm{DM}(\zi , R),	\label{eq:cosvar}
\end{equation}
where $b_\mathrm{g}$ and $\sigma_\mathrm{DM}(\zi , R)$ are the bias parameter of galaxies
and the density fluctuation of dark matter in a sphere with a radius $R$ at a redshift \zi, respectively.
We estimate $\sigma_\mathrm{DM}(\zi , R)$ with the growth factor, following \cite{Carroll92}
with the transfer function given by \cite{Bardeen86} \citep[see also][]{Mo02}.
The value of $\sigma_\mathrm{DM}(\zi , R)$ at $\zi = 2.2$ is estimated to be 0.055. 
Since \cite{Guaita10} find the bias parameter of $b_\mathrm{g} = 1.8 \pm 0.3$ from the clustering analysis of $\zi = 2.1$ LAEs
in the ECDF-S field, we adopt this value for $b_\mathrm{g}$ in Equation (\ref{eq:cosvar}).
We thus obtain the cosmic variance uncertainty of $\sigma_\mathrm{g}\simeq 0.099$.

\subsection{\lya\ Luminosity Functions} \label{sec:LFat2}

We derive the \lya\ LFs at $\zi = 2.2$ from the full
and EWgt60 samples, 
adopting the classical method of the \lya\ LF derivation 
\citep{Ouchi10,Konno14}
whose accuracy is confirmed by Monte-Carlo simulations
\citep{Shimasaku06,Ouchi08}.

We calculate \lya\ EWs of our LAEs from the aperture magnitudes of \textit{NB387} and \textit{B},
and obtain \lya\ luminosities of our LAEs from these EWs and the total magnitudes of \textit{NB387}.
We estimate photometric errors of \lya\ luminosities,
performing Monte-Carlo simulations
under the assumption that the SEDs of LAEs have a \lya\ line located at $\lambda_\mathrm{c}$ of \textit{NB387} and 
a flat UV continuum (i.e., $f_\nu = $ const.) with 
the inter-galactic medium (IGM) absorption of \cite{Madau95}.
We calculate volume number densities of LAEs in a \lya\ luminosity bin,
dividing the number counts of LAEs by our comoving survey volume 
($\simeq 1.32 \times 10^6$ Mpc$^3$; see Section \ref{sec:reduc})
under the assumption of the top-hat filter transmission curve.
We correct these number densities for the detection 
completeness estimated in Section \ref{sec:detcomp}.
Note that \cite{Ouchi08} investigate the incompleteness of the narrowband color selection based on the Monte-Carlo simulations,
and find that the incompleteness by the color is not significant.

\begin{figure}
\centering
\includegraphics[width=8.6cm]{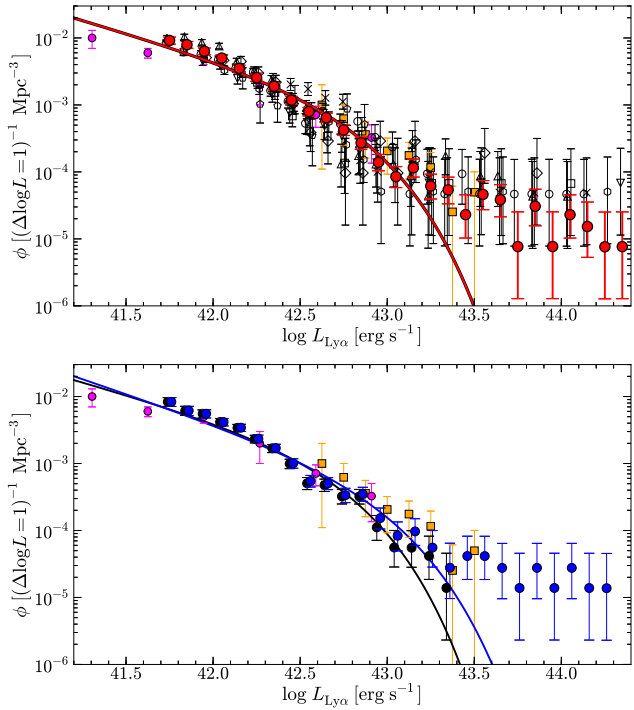}
\caption{\textit{Top}: \lya\ LF of our $\zi = 2.2$ LAEs with a luminosity bin of $\Delta \log L_{\mathrm{Ly}\alpha} = 0.1$.
The red filled circles represent the \lya\ LF derived 
from the full sample 
and the red solid curve denotes the best-fit Schechter function.
The black open symbols show the \lya\ LFs in 
the SXDS-C (squares), SXDS-N (diamonds), SXDS-S (hexagons), SXDS-W (pentagons),
COSMOS (circles), CDFS (inverted triangles), HDFN (triangles), and SSA22 (cross marks) fields.
For clarity, we slightly shift all the points along the abscissa.
The magenta filled circles and orange filled squares are the results from \cite{Cassata11} and \cite{Blanc11}, respectively.
\textit{Bottom}: \lya\ LF at $\zi = 2.2$ derived from the SXDS and COSMOS fields.
The blue and black filled circles represent the \lya\ LFs from the SXDS+COSMOS/All and SXDS+COSMOS/Culled subsamples, respectively.
The blue and black solid curves show the best-fit Schechter functions of our best-estimete \lya\ LFs
using the SXDS+COSMOS/All and SXDS+COSMOS/Culled subsamples, respectively.
The magenta filled circles and orange filled squares are the same as the top panel of this figure.
} 
\label{fig:LyaLF_hom30}
\end{figure}

The top panel of Figure \ref{fig:LyaLF_hom30} presents the best estimate of our \lya\ LF at $\zi =2.2$ from the full sample.
We also plot the \lya\ LF measurements derived from each-field data.
The error bars of the \lya\ LF include uncertainties from Poisson statistics and cosmic variance obtained in Section \ref{sec:cosvar}.
For the Poisson errors, we use the values in columns ``0.8413'' in Table 1 and 2 of \cite{Gehrels86} 
for the upper and lower limits of the Poisson errors, respectively.
The best estimate \lya\ LF covers a \lya\ luminosity 
range of $\log L_{\mathrm{Ly}\alpha} = 41.7-44.4$ erg s$^{-1}$.
Our \lya\ luminosity limit of
$\log L_{\mathrm{Ly}\alpha} = 41.7$ erg s$^{-1}$
($5.0 \times 10^{41}$ erg s$^{-1}$) 
is one order of magnitude fainter than the $L^{*}_{\mathrm{Ly}\alpha}$ values at $\zi = 3-6$
($\log L^{*}_{\mathrm{Ly}\alpha,\zi = 3-6} \sim 42.8$ erg s$^{-1}$;
\citealt{Shimasaku06, Gronwall07, Ouchi08}). 

\begin{figure*}
\centering
\includegraphics[width=\textwidth]{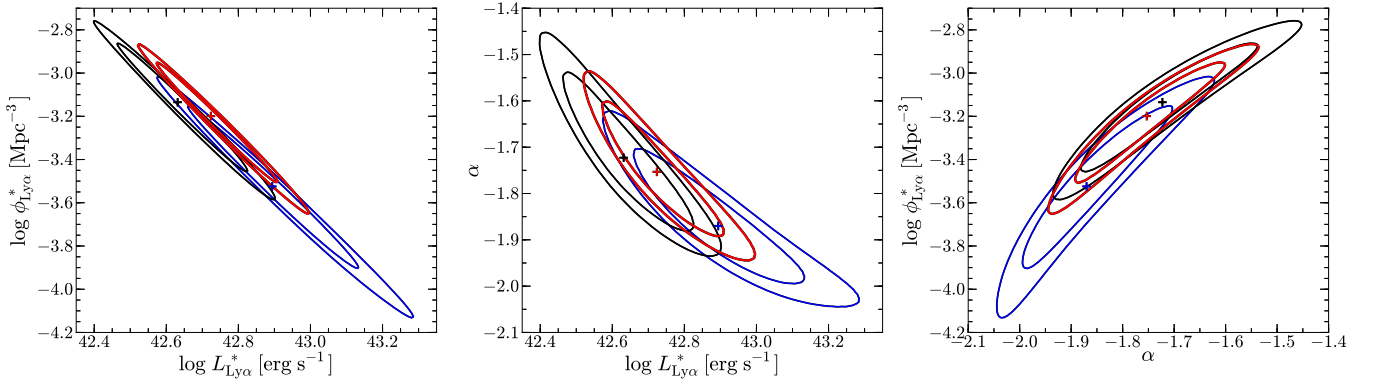}
\caption{Error contours of Schechter parameters, $L^{*}_{\mathrm{Ly}\alpha}$, $\phi^{*}_{\mathrm{Ly}\alpha}$ and $\alpha$.
The red contours represent our best-estimate \lya\ LF based on the full sample.
The blue and black contours show our best-estimate \lya\ LFs using the SXDS+COSMOS/All and SXDS+COSMOS/Culled subsamples, respectively.
The inner and outer contours denote the 68\% and 90\% confidence levels, respectively.
The red, blue and black crosses are the best-fit Schechter parameters
for our best-estimate \lya\ LFs based on the full sample, SXDS+COSMOS/All subsample, and SXDS+COSMOS/Culled subsample, respectively.
} 
\label{fig:LyaLF_hom30_cont}
\end{figure*}

We fit a Schechter function \citep{Schechter76} to our $\zi = 2.2$ \lya\ LF by minimum $\chi^2$ fitting.
The Schechter function is defined by
\begin{align}
&\phi_{\mathrm{Ly}\alpha} (L_{\mathrm{Ly}\alpha}) dL_{\mathrm{Ly}\alpha}	\notag	\\
= \ &\phi^{*}_{\mathrm{Ly}\alpha} \left( \frac{L_{\mathrm{Ly}\alpha}}{L^{*}_{\mathrm{Ly}\alpha}} \right) ^{\alpha} \exp \left( - \frac{L_{\mathrm{Ly}\alpha}}{L^{*}_{\mathrm{Ly}\alpha}} \right) d \left( \frac{L_{\mathrm{Ly}\alpha}}{L^{*}_{\mathrm{Ly}\alpha}} \right),
\end{align}
(see Section 1 for the definitions of the parameters).
For our fitting with the Schechter function, we use \lya\ LF measurements from the studies of
ours, \cite{Blanc11}, and \cite{Cassata11}.
We do not include the results from the other studies,
because there exist unknown systematics that is discussed in Section \ref{sec:compaLyaLF}.
We determine three parameters of the Schechter function simultaneously, 
and obtain the best-fit Schechter parameters of $\alpha= -1.75^{+0.10}_{-0.09}$,
$L^{*}_{\mathrm{Ly}\alpha} = 5.29^{+1.67}_{-1.13} \times 10^{42} \ \mathrm{erg} \ \mathrm{s}^{-1}$ and
$\phi^{*}_{\mathrm{Ly}\alpha} = 6.32^{+3.08}_{-2.31} \times 10^{-4} \ \mathrm{Mpc}^{-3}$.
This is the first time to determine three Schechter function parameters with
no fixed parameter(s), and the faint-end slope of $\alpha$ is reasonably well constrained.
Table \ref{table:param_total_cull} presents these best-fit Schechter parameters.
We show the best-fit Schechter function in the top panel of Figure \ref{fig:LyaLF_hom30},
and error contours of the Schechter parameters in Figure \ref{fig:LyaLF_hom30_cont}.

The top panel of Figure \ref{fig:LyaLF_hom30} shows an excess of the number densities beyond
the best-fit Schechter function at 
the bright-end of $\log L_{\mathrm{Ly}\alpha} \gtrsim 43.4$ erg s$^{-1}$.
We refer to this excess as bright-end hump.
In our Schechter function fit, we include the data of the bright-end hump.
Because the errors of the \lya\ LF at the faint end are significantly smaller than those at the bright end,
the best-fit parameters are not significantly changed by the inclusion of the bright-end hump data (see footnote of Table \ref{table:param_total_cull}).

\cite{Ouchi08} find that there is a possible excess of the \lya\ LFs at $\zi = 3.1$ and $3.7$
similar to the bright-end hump, and claim that
100\% of LAEs host AGNs at the bright ends of
$\log L_{\mathrm{Ly}\alpha} > 43.6$ and $43.4$ erg s$^{-1}$, respectively, 
based on the large-area LAE survey with the multiwavelength
data set.
Thus, the bright-end hump of our $\zi = 2.2$ \lya\ LF may be produced by
AGNs. To examine whether our LAEs at the bright end include AGNs,
we use the multiwavelength data of X-ray, UV, and radio available in the SXDS, COSMOS, CDFS,
HDFN, and SSA22 fields.
For the X-ray data, we use the \textit{XMM-Newton} source catalog in the SXDS field \citep{Ueda08},
the \textit{Chandra} 1.8 Ms catalog in the COSMOS field \citep{Elvis09},
the \textit{Chandra} 4 Ms source catalog in the CDFS field \citep{Xue11},
and the \textit{Chandra} 2 Ms catalog in the HDFN field \citep{Alexander03}.
The typical sensitivity limits of these X-ray data are $\sim 10^{-16}$ - $10^{-15}$ erg cm $^{-2}$ s$^{-1}$ for the SXDS and COSMOS fields,
and $\sim 10^{-17}$ - $10^{-16}$ erg cm $^{-2}$ s$^{-1}$ for the CDFS and HDFN fields.
We use \textit{GALEX} FUV and NUV images for the UV data,
and obtain these images from the Multimission Archive at STScI
(see also \citealt{Zamojski07} for the COSMOS field).
The \textit{GALEX} images reach the $3\sigma$ detection limit of $\sim 25$ - $26$ mag.
The Very Large Array 1.4 GHz source catalogs of \cite{Simpson06} (SXDS),
\cite{Schinnerer07} (COSMOS), and \cite{Miller13} (CDFS) are used for the radio data.
These radio data reach an rms noise level of $\sim 10$ $\mu$Jy beam$^{-1}$.
We find that a majority of our bright LAEs are detected in the
multiwavelength data, and summarize the numbers of these LAEs
in Table \ref{table:sample}.
Under the column of ``culled sample'' in Table \ref{table:sample} ,
we show the numbers of LAEs with no counterpart detection(s) in the X-ray, UV, and radio data.
As shown in Table \ref{table:sample}, the SXDS and COSMOS fields have the data 
that cover all of the X-ray, UV, and radio wavelengths.
Moreover, the X-ray, UV, and radio data spatially cover 
the entire fields of SXDS and COSMOS with the similar
sensitivities.
We make a subsample that is composed of 
all 1,576 LAEs 
found in the SXDS and COSMOS fields, and
refer to this subsample as SXDS+COSMOS/All.
We then make another subsample consisting of
1,538 LAEs with no multiwavelength counterpart detection(s) 
in the SXDS and COSMOS fields,
which is dubbed SXDS+COSMOS/Culled.

In the bottom panel of Figure \ref{fig:LyaLF_hom30},
we plot the \lya\ LFs derived from the subsamples of SXDS+COSMOS/All and SXDS+COSMOS/Culled.
We fit the Schechter function to these \lya\ LFs
and the complementary \lya\ LF data of \cite{Blanc11} and \cite{Cassata11}, 
and presents the 
best-fit 
Schechter
parameter sets 
and the error contours
in Table \ref{table:param_total_cull}
and Figure \ref{fig:LyaLF_hom30_cont}, respectively.
Comparing the \lya\ LF of the SXDS+COSMOS/All subsample with 
that of the full sample,
in Figures \ref{fig:LyaLF_hom30} and \ref{fig:LyaLF_hom30_cont},
we find that the \lya\ LF of the SXDS+COSMOS/All subsample is consistent with 
that of the full sample
within the uncertainties.
Figure \ref{fig:LyaLF_hom30_cont} indicates that 
the Schechter fitting results of the full sample and the SXDS+COSMOS/All subsample
are very similar with the one of the SXDS+COSMOS/Culled subsample, which are
determined in the wide luminosity range of $\log L_{\rm Ly\alpha} = 41.7-44.4$ erg s$^{-1}$.
However, there are no objects in SXDS+COSMOS/Culled subsample that has 
$\log L_{\mathrm{Ly}\alpha} > 43.4$ erg s$^{-1}$.
The \lya\ LF of SXDS+COSMOS/Culled subsample does not have a bright-end hump  
such found in those of the full sample and the SXDS+COSMOS/All subsample.
These comparisons suggest that the bright-end hump of the $\zi = 2.2$ \lya\ LF is originated
from AGNs that are bright in the X-ray, UV, and/or radio wavelength(s).
We discuss more details of the bright-end hump 
in Sections \ref{sec:excess} and \ref{sec:AGN_LF}. 

\input{param_total_cull}

\subsection{Comparison with Previous Studies} \label{sec:compaLyaLF}

\begin{figure*}
\centering
\includegraphics[width=\textwidth]{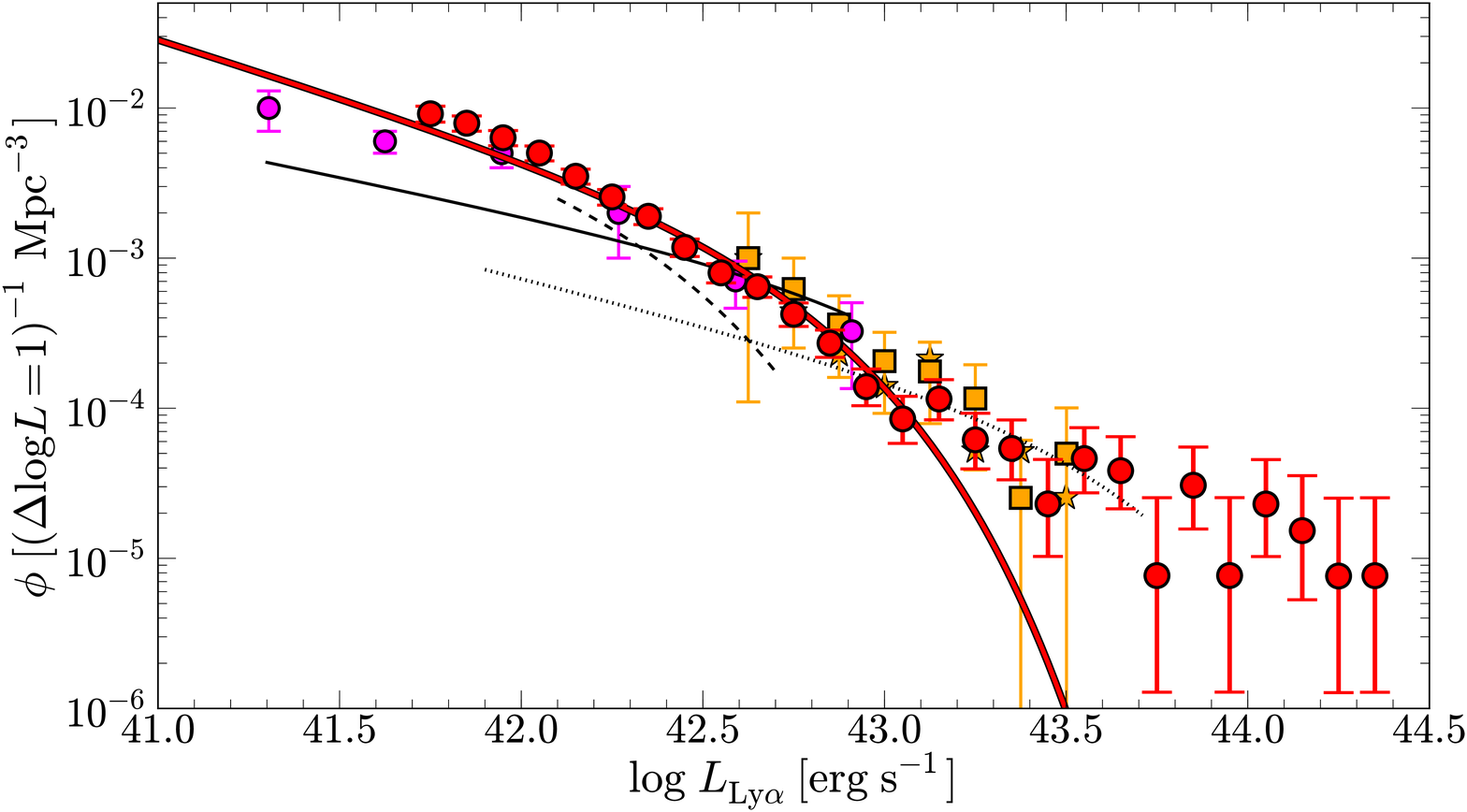}
\caption{Comparison of our $\zi = 2.2$ \lya\ LF with the previous measurements of \lya\ LF at $\zi \sim 2$.
The red filled circles denote our \lya\ LF and the red solid curve is the best-fit Schechter function,
which are the
same as the top panel of Figure \ref{fig:LyaLF_hom30}.
The magenta filled circles represent the \lya\ LF given by \cite{Cassata11} at $2 < \zi  < 3.2$.
The orange stars and squares show the LFs by \cite{Blanc11} based on the spectroscopic surveys of LAEs at
$1.9 < \zi < 2.8$ and $1.9 < \zi  < 3.8$, respectively.
The black solid, dashed, and dotted lines are the best-fit Schechter functions
obtained by \cite{Hayes10}, \cite{Ciardullo12}, and \cite{Ciardullo14}, respectively.
Since the previous \lya\ LF estimates are limited in the ranges of 
$\log L_{\mathrm{Ly}\alpha} = 41.3-42.9$ erg s$^{-1}$ \citep{Hayes10}, 
$42.1-42.7$ erg s$^{-1}$ \citep{Ciardullo12}, 
and $41.9-43.7$ erg s$^{-1}$ \citep{Ciardullo14}, 
we show the black lines within these ranges.  
} 
\label{fig:LyaLF_comp}
\end{figure*}

We compare our best-estimate \lya\ LF with those from previous studies at $\zi \sim 2$.
In Figure \ref{fig:LyaLF_comp}, we plot the \lya\ LFs obtained by narrowband imaging surveys 
(\citealt{Hayes10, Ciardullo12}; see also \citealt{Guaita10})
and blank-field spectroscopic surveys (\citealt{Blanc11, Cassata11, Ciardullo14}).
\cite{Hayes10} carry out deep imaging with two narrowband filters covering \lya\ and H$\alpha$ lines,
and report the \lya\ LF as well as the \lya\ escape fraction of $\zi = 2.2$ LAEs.
\cite{Ciardullo12} derive the \lya\ LF of $\zi = 2.1$ LAEs based on the narrowband data of \cite{Guaita10}. 
In both studies, the \lya\ EW criterion of narrowband excess colors is EW$_0 = 20\mathrm{\AA}$, 
comparable to our studies.
\cite{Blanc11} and \cite{Ciardullo14} obtain the \lya\ LFs by the spectroscopic observations of 
the Hobby Eberly Telescope Dark Energy Experiment (HETDEX) Pilot Survey
for LAEs at $1.9 < \zi < 3.8$ and $1.90 < \zi < 2.35$, respectively.
\cite{Cassata11} make a spectroscopic sample of LAEs at $2 < \zi < 6.6$ with the VIMOS VLT Deep Survey.
In these spectroscopic surveys, most of LAEs have a \lya\ EW greater than $20 \mathrm{\AA}$.
Table \ref{table:previous_studies} summarizes 
the best-fit Schechter parameters (and \lya\ luminosity ranges of the observations)
given by our and the previous studies.

In Figure \ref{fig:LyaLF_comp} and Table \ref{table:previous_studies},
we find that our $\zi = 2.2$ \lya\ LF is generally consistent with those of the previous studies
in the measurement ranges of the \lya\ luminosity overlaps.
However, there exist some noticeable differences. 
The \lya\ LF of \cite{Ciardullo12} is not similar to ours and \citet{Blanc11} at the bright end,
but similar to
ours and \citet{Cassata11} at the faint end. In contrast, the \lya\ LF of \cite{Ciardullo14} 
is not consistent with ours and \citet{Cassata11} at the faint end, but 
consistent with
ours and \citet{Blanc11} at the bright end.
Because \cite{Ciardullo12} and \cite{Ciardullo14} cover the reasonably wide
\lya\ luminosity ranges of $42.1 < \log L_{\rm Ly\alpha} < 42.7 $ erg s$^{-1}$
and $41.9 < \log L_{\rm Ly\alpha} < 43.7 $ erg s$^{-1}$, respectively,
the origins of these differences at the bright and faint ends are not clear.

As clarified in Table \ref{table:previous_studies}, most of the previous studies fit
the Schechter function to their \lya\ LFs, assuming 
a fixed parameter.
\cite{Hayes10} constrain three Schechter parameters simultaneously,
but the uncertainties of these parameters are large due to small statistics
(see also \citealt{Gronwall07} for $\zi \sim 3$).
Our study constrains three Schechter parameters simultaneously, using
the large LAE sample of 3,137 LAEs covering 
the wide \lya\ luminosity range ($\log L_{\rm Ly\alpha} = 41.7 - 44.4$ erg s$^{-1}$).

\input{previous_studies}

%%%%%%%%%%%%%%%%%%%%%%%%%%%%%%%%%%%%%%%%%%%%%%%%%%%%%%%%%%%%%%%%
\section{\lya\ Luminosity Function and\\
Density Evolution} \label{sec:evoLya}

\subsection{Evolution of \lya\ LFs} \label{sec:evoLF}

In this section, we first examine the evolution of \lya\ LFs at $\zi \sim 2 - 3$
and then investigate the evolution from $\zi \sim 0$ to $6$ 
with the compilation of the \lya\ LF data taken from the 
literature.

\begin{figure*}
\centering
\includegraphics[width=\textwidth]{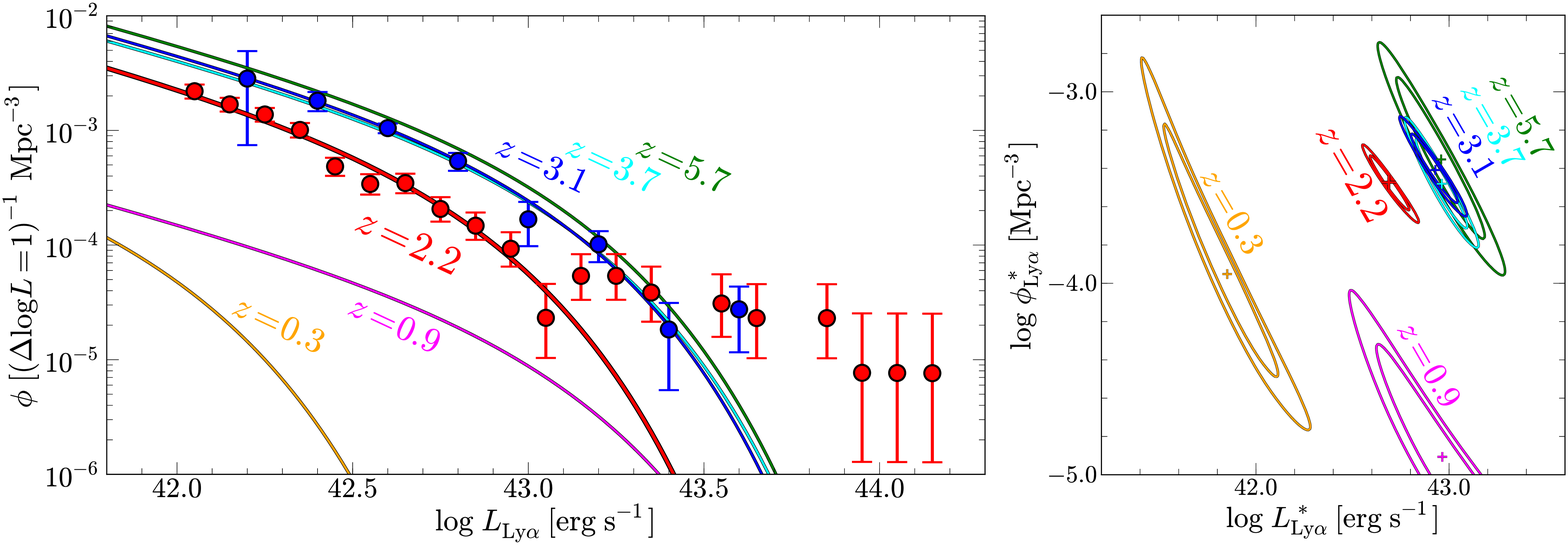}
\caption{\textit{Left}: Evolution of \lya\ LF 
from $z=0$ to $6$.
The red filled circles are our $\zi = 2.2$ \lya\ LF of
the EWgt60 sample, and the blue filled circles denote the LF at $\zi = 3.1$ derived by \cite{Ouchi08}.
The orange, magenta, red, blue, cyan, and green curves show the best-fit Schechter 
functions of the \lya\ LFs
at $\zi = 0.3$ \citep{Cowie10}, $0.9$ \citep{Barger12}, $2.2$ (this work), $3.1$, $3.7$, and $5.7$ \citep{Ouchi08}, respectively.
These Schechter functions are derived with a fixed slope value of $\alpha=-1.8$ 
that is the best-fit value of our $\zi = 2.2$ \lya\ LF.
\textit{Right}: Error contours of Schechter parameters, $L^{*}_{\mathrm{Ly}\alpha}$ and $\phi^{*}_{\mathrm{Ly}\alpha}$.
The orange, magenta, red, blue, cyan, and green contours represent 
the error contours of the Schechter parameters
at $\zi = 0.3$, $0.9$, $2.2$, $3.1$, $3.7$, and $5.7$, respectively.
The inner and outer contours indicate the $68$\% and $90$\% confidence levels, respectively.
}
\label{fig:LyaLF_cont}
\end{figure*}

For the $\zi \sim 3$ data, we use the \lya\ LF of \cite{Ouchi08}.
The $z=3.1$ \lya\ LF of \cite{Ouchi08} is derived in the same manner as ours 
(see Sections \ref{sec:detcomp}--\ref{sec:LFat2}).
Because the EW criterion of \cite{Ouchi08} is EW $\gtrsim 60$\AA ,
we compare the \lya\ LF obtained from our EWgt60 sample (Section \ref{sec:sample}).
The \lya\ LF and the best-fit Schechter function (parameters) for the EWgt60 sample
are presented in the left panel of Figure \ref{fig:LyaLF_cont} (Table \ref{table:schechter_evo}).
The left panel of Figure \ref{fig:LyaLF_cont} indicates that the \lya\ LFs increase from $\zi \sim 2$ to $3$.

To quantify this evolutionary trend, we show the error contours of the Schechter parameters of our $\zi =2.2$ \lya\ LF (red contours)
and the $\zi = 3.1$ \lya\ LF (blue contours) in the right panel of Figure \ref{fig:LyaLF_cont}.
Here, we apply our best-fit $\zi = 2.2$ \lya\ LF slope of $\alpha = -1.8$ (Section \ref{sec:LFat2}) 
to the $\zi = 3.1$ LF result, because $\alpha$ is not determined in the $\zi = 3.1$ \lya\ LF.
Comparing the $\zi = 2.2$ and $3.1$ error contours in the right panel of Figure \ref{fig:LyaLF_cont}, 
we find that the \lya\ LF increases from $\zi =2.2$ to $3.1$ at the $>90$\% confidence level.
However, this increase is not large, only within a factor of $\sim 2$ (see Table \ref{table:schechter_evo}).
Note that there exist no systematic errors raised by the analysis technique in the 
comparison of our $\zi=2.2$ and \citeauthor{Ouchi08}'s $z=3.1$ \lya\ LFs, because
our $\zi =2.2$ \lya\ LF is derived 
in the same manner as \citet{Ouchi08} based on the
similar Subaru narrowband data 
(Sections \ref{sec:detcomp}--\ref{sec:LFat2}).

We extend our investigation of \lya\ LF evolution from $\zi = 2 - 3$ to $\zi = 0 - 6$.
The left panel of Figure \ref{fig:LyaLF_cont} compares our best-estimate \lya\ LF at $\zi = 2.2$
with the \lya\ LFs at $\zi = 0.3$, $0.9$, $3.1$, $3.7$, and $5.7$ taken from the literature.
The right panel of Figure \ref{fig:LyaLF_cont} 
shows the error contours of our Schechter function fitting, where we fix
the $\alpha$ value to our best-fit slope $\alpha=-1.8$ of our $\zi = 2.2$ \lya\ LF.
The \lya\ LFs at $\zi = 0.3$ and $0.9$ are derived by the spectroscopic surveys 
with the \textit{GALEX} FUV and NUV grism data, respectively \citep{Cowie10, Barger12}.
We show the \lya\ LF measurements at $\zi = 3.7$ and $5.7$ given by \cite{Ouchi08}.
We summarize the best-fit Schechter parameters at $\zi = 0-6$ in Table \ref{table:schechter_evo}.
Note that EW$_0$ limits for the selection of LAEs are EW$_0$ $\gtrsim 10-30\mathrm{\AA}$ for all of the samples listed 
in Table \ref{table:schechter_evo} except for those of \citeauthor{Ouchi08}'s $\zi = 3.1$ and $3.7$ samples and our EWgt60 sample.
In the right panel of Figure \ref{fig:LyaLF_cont}, there is a significant increase of \lya\ LFs in $L^{*}_{\mathrm{Ly}\alpha}$ and/or $\phi^{*}_{\mathrm{Ly}\alpha}$
from $\zi \sim 0$ to $3$, albeit with the uncertain decrease of $\phi^{*}_{\mathrm{Ly}\alpha}$ from $\zi = 0.3$ to $0.9$,
which is first claimed by \cite{Deharveng08}.
The right panel of Figure \ref{fig:LyaLF_cont} also suggests no significant evolution of the \lya\ LFs at $\zi = 3-6$
that is concluded by \cite{Ouchi08}.

\input{schechter_evo}

\subsection{\lya\ Luminosity Density Evolution} \label{sec:evoLD}

We calculate the \lya\ luminosity densities (LDs), 
\begin{equation}
\rho^{\rm Ly \alpha}_{\rm obs} = \int^{\infty}_{L^{\rm Ly\alpha}_{\rm lim}} L_{\rm Ly\alpha} \phi_{\mathrm{Ly}\alpha} (L_{\mathrm{Ly}\alpha}) dL_{\mathrm{Ly}\alpha}, \label{eq:defLyaLD}
\end{equation}
at $\zi = 0 - 8$ with the \lya\ LFs shown in Section 4.1, 
where $L^{\rm Ly\alpha}_{\rm lim}$ is the \lya\ luminosity limit for the \lya\ LD estimates.
We choose the common \lya\ luminosity limit of $\log L^{\rm Ly\alpha}_{\rm lim} = 41.41$ erg s$^{-1}$
that corresponds to $0.03 L^{*}_{\mathrm{Ly}\alpha,\zi=3}$.

There are two systematic uncertainties for estimates of the \lya\ LDs.
One uncertainty is the choice of \lya\ luminosity limits.
The \lya\ luminosity limit can be lower than $\log L^{\rm Ly\alpha}_{\rm lim} = 41.41$ erg s$^{-1}$ 
to estimate representative \lya\ LDs. 
However, we confirm that 
the estimated \lya\ LDs are not largely different 
even if we integrate the \lya\ LFs down to a fainter luminosity of $\log L_{\mathrm{Ly}\alpha} = 40.0$ erg s$^{-1}$. 
The largest \lya\ LD difference of $\sim 0.4$ dex is found at $\zi = 0.3$, 
because the $L_{\rm Ly\alpha}^*$ 
value 
at $\zi = 0.3$ is significantly
smaller than those at the other redshifts.
Another uncertainty is \lya\ EW limits for selection of LAEs.
\lya\ LDs are based on LAE samples selected with 
a \lya\ EW limit
(i.e., EW$_0$ $\gtrsim 10-30$\AA).
\cite{Ouchi08} estimate \lya\ LDs for all (EW $> 0$\AA) LAEs 
and find that the \lya\ LDs are slightly larger 
than those for their EW-limited LAE samples (EW$_0$ $\gtrsim 10-30$\AA\ )
by $\sim 0.1$ dex at most.
These 
levels of differences do not change the results of the \lya\ LD evolution in this Section
that is at the level of an order of magnitude.
For these \lya\ LDs, we do not correct the \lya\ flux attenuation by neutral hydrogen (\HI) in the IGM.
The \lya\ LDs represent the amount of \lya\ photons escaping not only from ISM of galaxies,
but also from the \HI\ IGM.

\begin{figure*}
\centering
\includegraphics[width=0.8\textwidth]{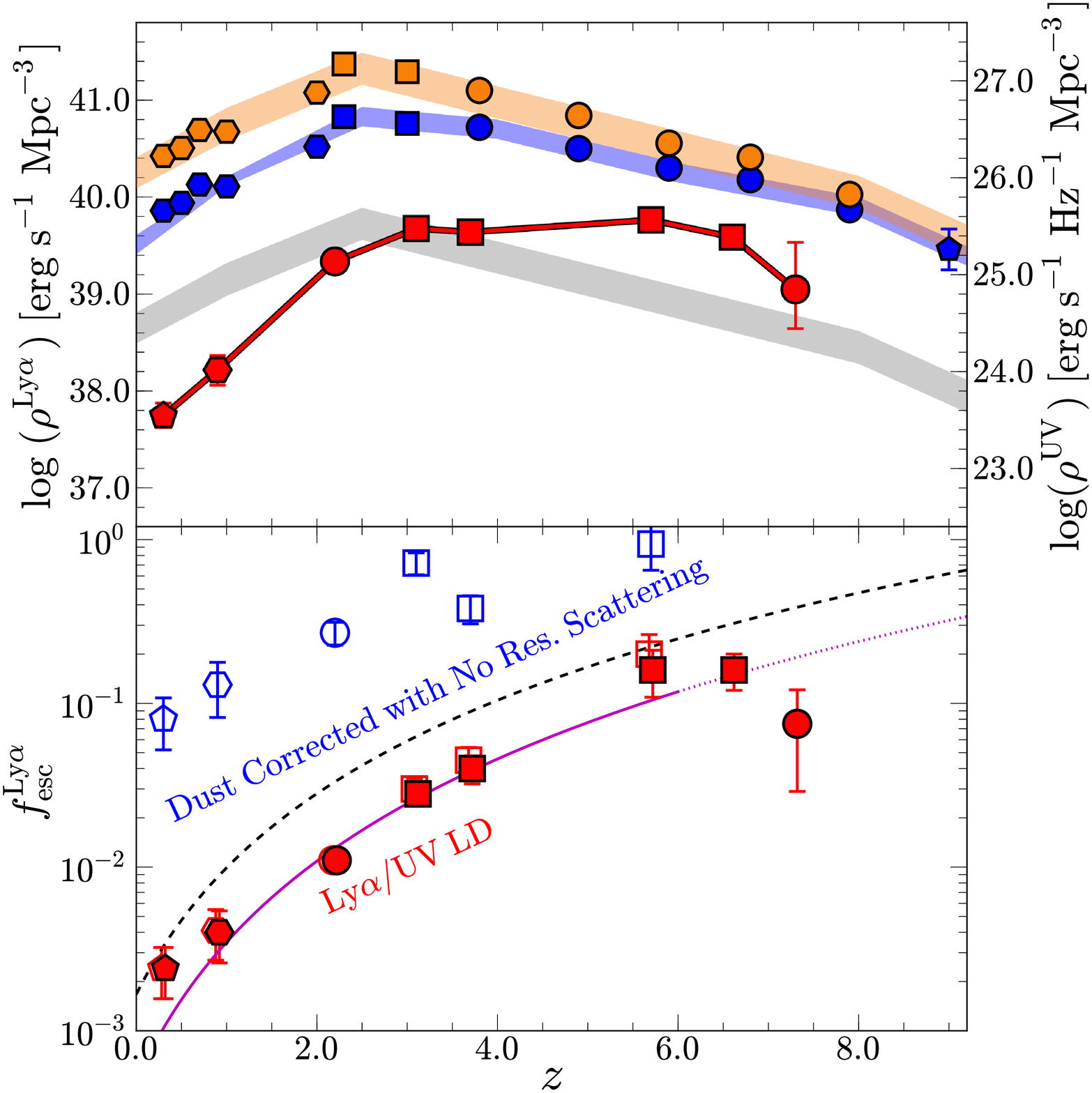}
\caption{\textit{Top}: Evolution of \lya\ LDs and 
UV LDs as a function of redshift.
The red circle at $\zi = 2.2$ shows the \lya\ LD obtained by this study.
The red pentagon at $\zi =0.3$ and hexagon at $\zi = 0.9$ are the \lya\ LDs derived by \cite{Cowie10} and \cite{Barger12}, respectively.
The red squares at $\zi = 3.1$, $3.7$, $5.7$, and $6.6$ denote the results of \cite{Ouchi08, Ouchi10},
and the red circle at $\zi = 7.3$ is the measurement given by \cite{Konno14}.
The blue symbols and shaded area represent the evolution of the dust-uncorrected UV LDs.
The blue pentagons at $\zi = 0-2$ and squares at $\zi = 2-3$ are the UV LDs obtained by \cite{Schiminovich05} and \cite{Reddy09}, respectively.
The blue circles and pentagon show the UV LDs given by \cite{Bouwens15} for $\zi=3.8$, $4.9$, $5.9$, $6.8$, and $7.9$,
and \cite{Ellis13} for $\zi = 9.0$, respectively.
The orange symbols and shaded area are 
the same as the blue ones, but for
the dust-corrected UV LDs.
The gray shaded area denotes the evolutionary tendency of the dust-corrected UV LDs scaled to the \lya\ LD at $\zi \sim 3$
for comparison.
\textit{Bottom}: Evolution of \lya\ escape fraction, $f^{\mathrm{Ly}\alpha}_{\mathrm{esc}}$, as a function of redshift.
The red filled symbols show the \lya\ escape fractions derived from the observed \lya\ LDs and dust-corrected UV LDs 
(Equation \ref{eq:fesc}).
The red open symbols represent our \lya\ escape fraction values corrected for IGM absorption using the relation of \cite{Madau95}.
The blue open symbols indicate the \lya\ escape fractions corrected for  
dust extinction in the case of no \lya\ resonance scattering (Equation \ref{eq:screen1}).
The magenta solid line is the best-fit function for our \lya\ escape fraction evolution 
from $z = 0$ to $6$
($f^{\mathrm{Ly}\alpha}_{\mathrm{esc}} = 5.0 \times 10^{-4} \times (1 + \zi)^{2.8}$),
while the black dashed line is the best-fit function 
derived by \cite{Hayes11}.
The magenta dotted line represents the extrapolation of the magenta solid line to $z>6$.
}
\label{fig:LyaLD_evo}
\end{figure*}

For comparison, we also use UV LDs taken from the literature \citep{Bouwens15}. The UV LD is defined by
\begin{equation}
\rho^{\rm UV}_{\rm obs} = \int^{\infty}_{L^{\rm UV}_{\rm lim}} L_{\rm UV} \phi_{\mathrm{UV}} (L_{\mathrm{UV}}) dL_{\mathrm{UV}},
\end{equation}
where $L^{\rm UV}_{\rm lim}$ is the UV luminosity limit for the UV LD estimates,
and $\phi_{\mathrm{UV}} (L_{\mathrm{UV}})$ is the best-fit Schechter function for the UV LF measurements.
Here, the value of $L^{\rm UV}_{\rm lim}$ is $0.03 L^{*}_{\mathrm{UV},\zi=3}$ ($M_{\rm UV}= -17.0$ mag).
The upper panel of Figure \ref{fig:LyaLD_evo} presents the evolution of the \lya\ LDs as a function of redshift
whose data are summarized in Table \ref{table:schechter_evo}. In the upper panel of Figure \ref{fig:LyaLD_evo},
we also plot the UV LDs of dust-uncorrected and -corrected UV LDs obtained by \cite{Bouwens15}.
Similar to the evolutionary trends of \lya\ LFs described in Section 4.1, 
we find the significant increase of \lya\ LDs from $\zi \sim 2$ to $3$ beyond the
measurement errors. Moreover, there is an rapid increase of \lya\ LDs by 
nearly two order of magnitudes from $\zi \sim 0$ to $3$,
and a plateau of \lya\ LDs between $\zi \sim 3$ and $6$.
The decrease of \lya\ LDs at $\zi \gtrsim 6$ is also found.
For more details, see Section 4.1 and the literature
(e.g. \citealt{Deharveng08, Ouchi08, Cowie10, Cowie11, Ciardullo12, Barger12, Wold14, Konno14}).

The \lya\ LD evolution is different from
the UV LD evolution in the upper panel of Figure \ref{fig:LyaLD_evo}.
There is an increase of UV LDs from $\zi \sim 0$ to $3$, but the
increase is only about an order of magnitude that is not as large
as the one of \lya\ LDs. At $\zi \sim 3-6$, the UV LDs show a moderate decrease 
and no evolutionary plateau like the one found in the \lya\ LD evolution.
At $\zi \gtrsim 6$, the decrease of \lya\ LDs is faster than the one of UV LDs
toward high-$\zi$.
We discuss the physical origins of these differences in Section \ref{sec:dust}.

%%%%%%%%%%%%%%%%%%%%%%%%%%%%%%%%%%%%%%%%%%%%%%%%%%%%%%%%%%%%%%%%
\section{Discussion} \label{sec:discuss}

\subsection{Bright-End Hump of the \lya\ LF} \label{sec:excess}
In the upper panel of Figure \ref{fig:LyaLF_hom30}, we find the bright-end hump of our $\zi = 2.2$ \lya\ LF
at $\log L_{\mathrm{Ly}\alpha} \gtrsim 43.4$ erg s$^{-1}$. The objects in the bright-end hump have
UV continuum magnitudes of $M_{\rm UV}\gtrsim -25$.
There are two possibilities to explain this hump.
One possibility is the existence of AGNs which have a strong \lya\ emission line (e.g., \citealt{Ouchi08}).
Another possibility is the magnification bias (e.g., \citealt{Wyithe11, Mason15}). 
The gravitational lensing of foreground massive galaxies 
increases
luminosities of LAEs at $\zi = 2.2$ that make the hump at the bright end LF.
The lower panel of Figure \ref{fig:LyaLF_hom30} shows that
all galaxies brighter than $\log L_{\mathrm{Ly}\alpha} = 43.4$ erg s$^{-1}$ have (a) bright counterpart(s) 
in X-ray, UV, and/or radio data, suggesting that these galaxies have AGNs.
If we remove these galaxies from our sample,
the shape of the \lya\ LF is explained by the simple 
Schechter function with no hump
(see the black solid line and black filled circles in the lower panel of Figure \ref{fig:LyaLF_hom30}).
These results indicate that the bright-end hump is almost fully explained by AGNs 
that have magnitudes of $M_{\rm UV}\gtrsim -25$.
These AGNs are significantly fainter than QSOs, and regarded as faint AGNs.
The magnification bias would exist, but it is very weak.
The major physical mechanism of the bright-end hump is not the magnification bias.

\subsection{Faint AGN UV LF} \label{sec:AGN_LF}

\begin{figure*}
\centering
\includegraphics[width=\textwidth]{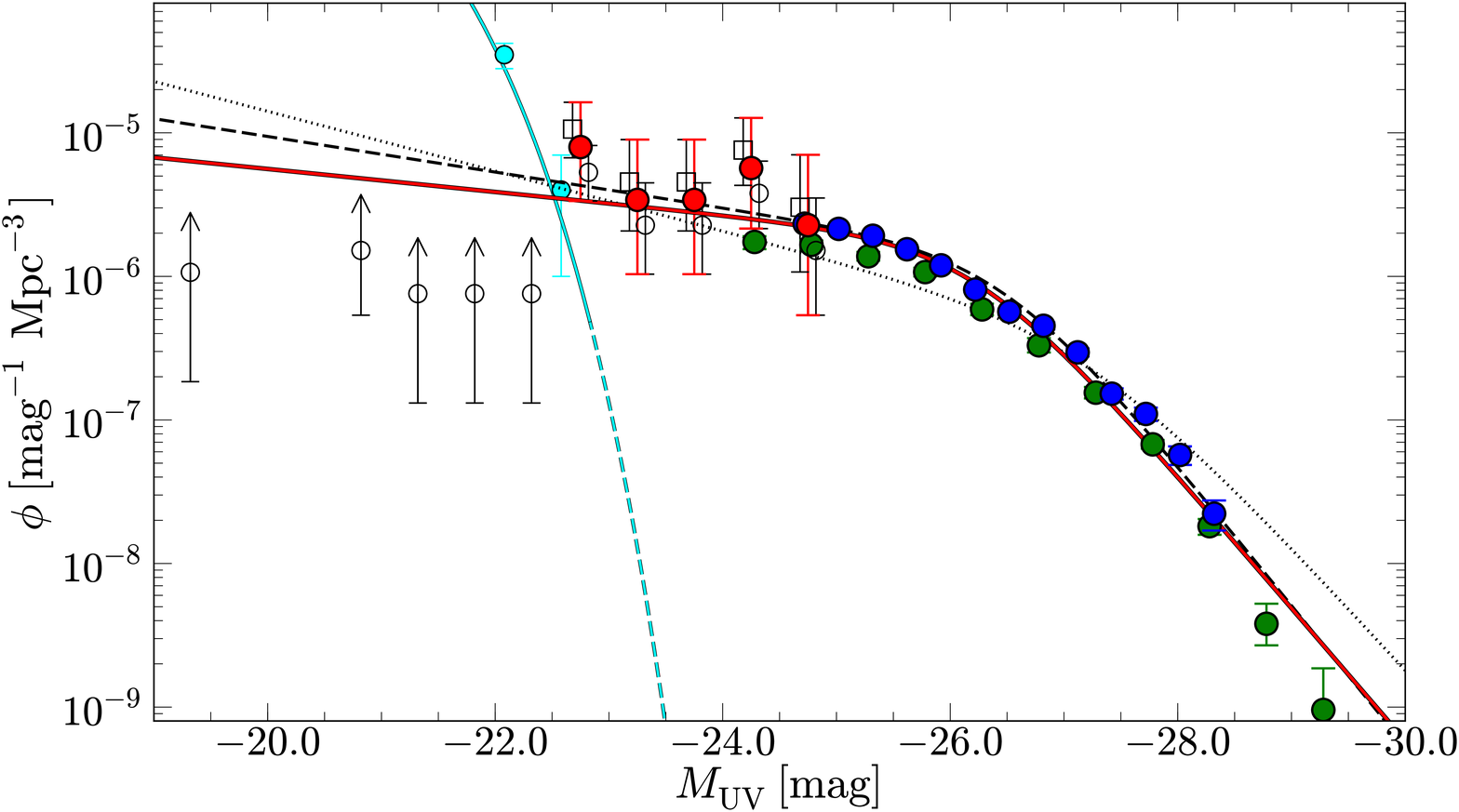}
\caption{UV LF of faint AGNs.
The red filled circles denote the best-estimate AGN UV LFs 
and the black open circles and squares represent the raw and maximally-corrected AGN UV LFs, 
respectively (see the text for details).
At $M_{\mathrm{UV}} \gtrsim -22.5$, we plot only the raw UV LF as lower limits with black arrows,
because one cannot estimate the incompleteness 
at this range (see the text for details).
For display purposes,
we slightly shift the black symbols along the abscissa.
The blue and green circles are the AGN UV LFs at $\zi \sim 2.2$ derived 
from the SDSS DR9 dataset \citep{Ross13}
and the 2dF-SDSS LRG and QSO survey dataset \citep{Croom09}, respectively.
The red curve shows the best-fit function for the AGN UV LFs of ours, \cite{Ross13}, and \cite{Croom09}.
The black dotted and dashed curves represent the best-fit functions
under the assumptions of the PLE and LEDE models introduced by \cite{Ross13}, respectively.
We also display the UV LF of $\zi = 2$ LBGs obtained by \cite{Reddy09} with the cyan circles.
The cyan solid curve represents the best-fit Schechter function of 
the LBG UV LF within a range of the observed UV-continuum magnitude (i.e., $M_\mathrm{UV} > -22.8$),
while the cyan dashed curve denotes the function extrapolated to $M_\mathrm{UV} < -22.8$.
} 
\label{fig:AGN_UVLF}
\end{figure*}

In Section \ref{sec:excess}, we discuss that the bright-end hump is made of faint AGNs 
($\log L_{\mathrm{Ly}\alpha} > 43.4$ erg s$^{-1}$),
all of which have the counterpart(s) in the 
X-ray, UV, and radio data.
Using the abundance and the UV continuum magnitudes ($M_\mathrm{UV} \gtrsim -25$) of these faint AGNs,
we derive faint AGN UV LFs.
These faint AGN UV LFs complement
the bright AGN UV LFs obtained by
cosmological large scale surveys such as Sloan Digital Sky Survey (SDSS).
To estimate the faint AGN UV LFs,
we measure \textit{i}-band magnitudes at the positions of the faint AGNs.
Here, we choose the \textit{i}-band magnitudes for UV continuum magnitude
estimates, because we compare our results with
the SDSS AGN study of \cite{Ross13} who use 
\textit{i}-band magnitudes to derive their AGN UV LF.
All of our faint AGNs are detected at the $> 5\sigma$ levels in our \textit{i}-band images.
Note that the $5\sigma$ limiting magnitudes of our \textit{i}-band images correspond to 
$M_{\rm UV}=-17.9$, $-18.6$, $-20.2$, $-19.7$, and $-18.5$ mag for the faint AGNs at $\zi = 2.2$
in the SXDS, COSMOS, CDFS, HDFN, and SSA22 fields, respectively.
We calculate the volume number densities of the faint AGNs in a UV-continuum magnitude bin,
dividing the number counts of faint AGNs by our comoving survey volume ($\simeq 1.32 \times 10^6$ Mpc$^3$).
Figure \ref{fig:AGN_UVLF} presents these UV LFs of our faint AGNs 
with black open circles that we call raw UV LFs.
The errors of the raw UV LFs are the Poisson errors for small number statistics \citep{Gehrels86}.

\begin{figure}
\centering
\includegraphics[width=8.0cm]{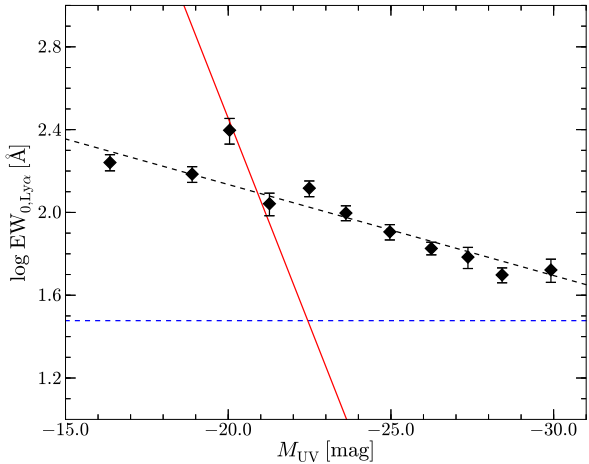}
\caption{\lya\ EW$_0$ as a function of UV-continuum magnitude of AGN.
The black diamonds represent the median values of the observed \lya\ EW$_0$ at a given UV-continuum magnitude,
and the black dashed line is a best-fit linear function obtained by \cite{Dietrich02}.
The error bars of the black diamonds indicate the measurement uncertainties of the \lya\ EWs.
The red solid line shows a locus of the luminosity for $\log L_{\mathrm{Ly}\alpha} = 43.4$ erg s$^{-1}$,
which is a selection criterion for our faint AGNs.
The blue dashed line denotes the EW$_0$ threshold for selection of our $\zi = 2.2$ LAEs (i.e., $\sim 20 - 30\mathrm{\AA}$).
} 
\label{fig:AGNcomp}
\end{figure}

Because AGNs do not always
have \lya\ emission that can be identified by our narrowband observations,
the raw UV LFs are incomplete. The raw UV LFs are regarded as the lower limits of the AGN UV LFs.
To evaluate the incompleteness, we use the relation of \lya\ EWs and UV-continuum magnitudes
(the Baldwin effect)
given by \cite{Dietrich02}.
\cite{Dietrich02} obtain the median values of \lya\ EWs at a given UV-continuum magnitude bin
based on 744 AGNs at $\zi \sim 0-5$, where a negligibly small fraction ($\sim 10$\%) of 
damped Ly$\alpha$ systems and low quality data is removed from their AGN sample.
Note that we do not take into account UV continuum indices of AGNs,
because our sample is too small to make statistically useful subsamples with the additional parameter of the UV continuum indices.
In Figure \ref{fig:AGNcomp},
we plot the median values with 
the black filled diamonds. 
Because no PDFs of \lya\ EWs are presented in \cite{Dietrich02},
the errors of the black filled diamonds represent
the measurement uncertainties of \lya\ EWs.
Figure \ref{fig:AGNcomp} shows a correlation, 
indicating that UV-continuum faint AGNs have
large \lya\ EWs. 
The red and blue lines in Figure \ref{fig:AGNcomp} represent
our selection limits of $\log L_{\mathrm{Ly}\alpha} > 43.4$ erg s$^{-1}$ (for the objects in
the bright-end hump) and the EW$_0\gtrsim 20 - 30$\AA\ (for our LAE sample), respectively.
In Figure \ref{fig:AGNcomp}, we find that these selection limits (red and blue lines) 
are far below the median values (black diamonds) at $M_{\rm UV}\lesssim -22.5$.
Thus, the faint AGN UV LFs at $M_{\rm UV}\lesssim -22.5$ 
can be determined
with reasonable completeness corrections.
Because the \lya\ EW PDFs are not given in \cite{Dietrich02},
one cannot simply estimate the incompleteness.
However, all of the median values at $M_{\rm UV} \lesssim -22.5$
are placed above the selection limits. The maximum correction factor is $\sim 2$
in the most extreme case that the \lya\ EW PDF has the bottom heavy distribution.
This is because about a half of the AGNs at maximum could fall below our selection limits, 
which can keep the median values as high as those obtained by \cite{Dietrich02}.
For our faint AGNs at $M_{\rm UV} \lesssim -22.5$,
we correct the raw UV LFs for the incompleteness with the maximum correction factor,
and plot the maximally-corrected UV LFs with the open squares in Figure \ref{fig:AGN_UVLF}.
Because the real UV LFs should be placed between 
the raw UV LFs and the maximally-corrected UV LFs,
we define the best-estimate UV LFs by the average of the raw and maximally-corrected UV LFs
with the conservative error bars that completely cover the $1\sigma$ uncertainties of these two UV LFs.
The red circles in Figure \ref{fig:AGN_UVLF} represent 
the best-estimate UV LFs.
In Figure \ref{fig:AGN_UVLF} , 
we also present the AGN UV LFs at $\zi \sim 2.2$ derived with the SDSS DR9 data (the blue circles; \citealt{Ross13})
and the 2dF-SDSS LRG and QSO survey data (the green circles; \citealt{Croom09}).
There is a magnitude-range overlap of our, \citeauthor{Ross13}'s, and \citeauthor{Croom09}'s AGN UV LF estimates at $M_\mathrm{UV} \simeq -24.8$.
The number densities from our, \citeauthor{Ross13}'s, and \citeauthor{Croom09}'s studies agree very well within the uncertainties at the overlap magnitude,
indicating that our AGN UV LF estimates are reliable.
We also confirm that the AGN UV LF in our study is also consistent with that in \cite{Jiang06}.

We fit a double power-law function to the AGN UV LFs of ours, \cite{Ross13}, and \cite{Croom09}.
The double power-law function for the AGN number density, $\phi_{\rm AGN} (M_\mathrm{UV})$, 
is defined by
\begin{align}
&\phi_{\rm AGN} (M_\mathrm{UV})	\notag	\\
= \ &\frac{ \phi_{\rm AGN}^{*} }{ 10^{0.4(\alpha_{\rm AGN} + 1)(M_\mathrm{UV} - M_{\rm AGN}^{*})} + 10^{0.4(\beta_{\rm AGN} + 1)(M_\mathrm{UV} - M_{\rm AGN}^{*})}},
\end{align}
where $\phi_{\rm AGN}^{*}$ and $M_{\rm AGN}^{*}$ are the characteristic number density and magnitude of AGNs, respectively.
The parameters of $\alpha_{\rm AGN}$ and $\beta_{\rm AGN}$ determine
the faint- and bright-end slopes of the AGN UV LFs.
We obtain the best-fit parameters of 
$\phi_{\rm AGN}^{*} = 1.8 \pm 0.2 \times 10^{-6}$ Mpc$^{-3}$, 
$M_{\rm AGN}^{*} = -26.2 \pm 0.1$,
$\alpha_{\rm AGN} = -1.2 \pm 0.1$, and 
$\beta_{\rm AGN} = -3.3 \pm 0.1$,
and present the best-fit function with the red line
in Figure \ref{fig:AGN_UVLF}.
Our results suggest that the faint-end slope $\alpha_{\rm AGN}$ 
is moderately flat at $M_{\rm UV}\simeq -23$ - $-25$.

\cite{Ross13} and \cite{Croom09} show the faint-end slopes at $\zi \sim 2.2$ are $\alpha_{\rm AGN} =  -1.3^{+0.7}_{-0.1}$
and $-1.4 \pm 0.2$, respectively, that are consistent with our result. 
Because relatively steep faint-end slopes ($\alpha_{\rm AGN} \simeq -1.5$ - $-1.8$) are
obtained for $\zi = 4 - 6.5$ AGNs \citep{Ikeda11,Giallongo15},
our moderately flat faint-end slope at $\zi \sim 2.2$ would suggest
that the faint-end slope steepens toward high-$\zi$.
Figure \ref{fig:AGN_UVLF} displays the two models of a pure luminosity evolution (PLE) model 
and a luminosity evolution and density evolution (LEDE) model that are introduced by \cite{Ross13}.
Comparing these two models, 
we find that the LEDE model explains our AGN UV LFs better than the PLE model.
This comparison suggests that the AGN UV LF evolution 
involves both luminosities and densities.

\subsection{\lya\ Escape Fraction Evolution and \\
the Physical Origins} \label{sec:dust}

In Section \ref{sec:evoLD}, we compare the evolution of the \lya\ and UV LDs,
and conclude that
the evolutions of \lya\ and UV LDs are different.
To understand the physical origins of the differences between \lya\ and UV LD evolutions, 
we investigate evolution of \lya\ escape fractions, $f^{\mathrm{Ly}\alpha}_{\mathrm{esc}}$.
The \lya\ escape fraction evolution is investigated 
by previous studies (e.g., \citealt{Hayes11, Blanc11}).
In this study, we revisit the \lya\ escape fraction evolution,
because there are significant progresses on the estimates of \lya\ LDs 
from recent Subaru, VLT, and HETDEX pilot surveys (e.g., \citealt{Cassata11,Ciardullo14,Konno14})
and UV LDs from HST UDF12, CANDELS, and HFF programs (e.g., \citealt{Bouwens15}).

The \lya\ escape fraction is defined by
\begin{equation}
f^{\mathrm{Ly}\alpha}_{\mathrm{esc}} = \rho^{\mathrm{obs}, \mathrm{Ly}\alpha}_{\mathrm{SFRD}} / \rho^{\mathrm{int}, \mathrm{UV}}_{\mathrm{SFRD}},	\label{eq:fesc}
\end{equation}
where $\rho^{\mathrm{obs}, \mathrm{Ly}\alpha}_{\mathrm{SFRD}}$ is
the star formation rate densities (SFRDs) estimated from the observed \lya\ LDs.
The variable of $\rho^{\mathrm{int}, \mathrm{UV}}_{\mathrm{SFRD}}$ represents 
SFRDs calculated from the intrinsic UV LDs that are UV LDs corrected for dust extinction.
Note that the contribution from AGN luminosities to \lya\ LDs and UV LDs are negligibly small
due to the low AGN abundance, and that we regard these \lya\ and UV photons are produced by
star formation.

We use the \lya\ LDs shown in Figure \ref{fig:LyaLD_evo} (Section \ref{sec:evoLD}),
and derive $\rho^{\mathrm{obs}, \mathrm{Ly}\alpha}_{\mathrm{SFRD}}$.
In the estimation of star-formation rates (SFRs) from the \lya\ luminosities,
we apply
\begin{equation}
\mathrm{SFR} \ (M_\odot \ \mathrm{yr}^{-1}) = L_{\mathrm{Ly}\alpha} \ (\mathrm{erg} \ \mathrm{s}^{-1})/(1.1 \times 10^{42}),
\label{eq:sfr_lya}
\end{equation}
that is the combination of the H$\alpha$ luminosity-SFR relation \citep{Kennicutt98} 
and the case B approximation \citep{Brocklehurst71}.
For $\rho^{\mathrm{int}, \mathrm{UV}}_{\mathrm{SFRD}}$ values,
we use the dust-extinction corrected 
SFRDs 
derived by \citet{Bouwens15}. The SFRDs are estimated from the UV LDs
that are integrated values of UV LFs down to $0.03 L^{*}_{\mathrm{UV},\zi = 3}$ (Section 4.2).
The SFRs are estimated from UV luminosities with the equation \citep{Madau98},
\begin{equation}
\mathrm{SFR} \ (M_\odot \ \mathrm{yr}^{-1}) = L_\mathrm{UV} \ (\mathrm{erg} \ \mathrm{s}^{-1} \ \mathrm{Hz}^{-1})/(8 \times 10^{27}),
\label{eq:sfr_uv}
\end{equation}
where $L_\mathrm{UV}$ is the UV luminosity measured at 1500 \AA.
The dust extinction values are evaluated from the UV-continuum slope measurements
with the relation of \cite{Meurer99}.
The UV LDs corresponding to these SFRDs are presented in Figure \ref{fig:LyaLD_evo}.
Note that the Salpeter IMF is assumed in Equations (\ref{eq:sfr_lya}) and (\ref{eq:sfr_uv}). 

From these SFRDs, we estimate \lya\ escape fractions
with Equation (\ref{eq:fesc}).
The bottom panel of Figure \ref{fig:LyaLD_evo}
presents the \lya\ escape fractions at $\zi \sim 0 - 8$.
We fit a power-law function of $\propto (1+\zi)^n$ to these \lya\ escape fraction estimates at $\zi \sim 0 - 6$,
where $n$ is the power law index.
We obtain the best-fit function of 
$f^{\mathrm{Ly}\alpha}_{\mathrm{esc}} = 5.0\times 10^{-4} \times (1 + \zi)^{2.8}$.
The best-fit function is shown in the bottom panel of Figure \ref{fig:LyaLD_evo}.
The best-fit function indicates a large increase of \lya\ escape fractions
from $\zi \sim 0$ to $6$ by two orders of magnitude, 
although the data points of $\zi \gtrsim 6$ depart from the best-fit function.
This trend is similar to the one claimed by \citet{Hayes11}. We compare
the results of \citet{Hayes11} with this study in the bottom panel of Figure \ref{fig:LyaLD_evo}.
Although the general evolutionary trend is the same in \citeauthor{Hayes11}'s and our results, there is
an offset between these two results.
This offset is explained by the differences of the \lya\ and UV luminosity limits
for deriving the \lya\ and UV LDs from \lya\ and UV LFs, respectively.
In fact, we obtain \lya\ escape fractions consistent with those of \cite{Hayes11},
if we calculate the \lya\ escape fractions with the \lya\ and UV luminosity limits 
same as those of \cite{Hayes11}. In other words, the choice of \lya\ and UV luminosity limits
moderately change the \lya\ escape fraction estimates, but 
the two-orders of magnitude evolution of \lya\ escape fractions
is significantly larger than these changes.
It should be noted that, if we calculate $f^{\mathrm{Ly}\alpha}_{\mathrm{esc}}$ with our \lya\ LF and
the \citeauthor{Sobral13}'s H$\alpha$ LFs, we obtain $f^{\mathrm{Ly}\alpha}_{\mathrm{esc}}=0.013$
that is consistent with our original estimate with the UV LFs ($f^{\mathrm{Ly}\alpha}_{\mathrm{esc}}=0.011$).
Thus, there are no significant systematics in $f^{\mathrm{Ly}\alpha}_{\mathrm{esc}}$ estimates
for the choices of UV and H$\alpha$ LFs.
Recently, \cite{Matthee16} obtain the $f^{\mathrm{Ly}\alpha}_{\mathrm{esc}}$ value at $\zi = 2.2$ from the \lya /H$\alpha$ flux measurements
of their 17 H$\alpha$ emitters.
They obtain a median value of $f^{\mathrm{Ly}\alpha}_{\mathrm{esc}} = 0.016\pm0.005$ that is also consistent with ours.

At $\zi \gtrsim 6$, there exist the departures of the \lya\ escape fraction estimates
from the best-fit function (the bottom panel of Figure \ref{fig:LyaLD_evo}). Moreover, the departure becomes larger toward high-$\zi$.
There is a decrease of \lya\ escape fractions from $\zi \sim 6$ to $8$ by a factor of $\sim 2$.
Because the redshift range of $\zi \gtrsim 6$ corresponds to the epoch of reionization (EoR),
this decrease of \lya\ escape fractions at $\zi \gtrsim 6$ is explained by the increase of
\lya\ scattering of \HI\ in the IGM at the EoR. 
In other words, it is likely that the physical origin of the $f^{\mathrm{Ly}\alpha}_{\mathrm{esc}}$ decrease at $\zi \gtrsim 6$ is cosmic reionization.
This result is in the different form of the previous results that claim the signature of
cosmic reionization based on the \lya\ luminosity function decrease at $\zi > 6$
(e.g. \citealt{Kashikawa06, Ouchi10, Kashikawa11, Shibuya12, Jiang13, Konno14})
and the \lya-emitting galaxy fraction decrease at $\zi > 6$ (e.g. \citealt{Pentericci11, Ono12, Treu13, Schenker14}).

Here, we discuss the physical mechanism of the large, 
two orders of magnitude
increase of $f^{\mathrm{Ly}\alpha}_{\mathrm{esc}}$ from $\zi \sim 0$ to $6$. 
Note that $f^{\mathrm{Ly}\alpha}_{\mathrm{esc}}$ is defined as 
the ratio of the \lya\ LD to the UV LD of star-forming galaxies. 
Since these LDs are mainly contributed 
by continuum faint galaxies with $M_{\rm UV}\gtrsim -19$,
majority of which
show \lya\ in emission
(\citealt{Stark10}), 
we regard LAEs as a dominant population of 
high-$z$ star-forming galaxies 
in the following discussion. 

There are four
possible physical mechanisms for the large $f^{\mathrm{Ly}\alpha}_{\mathrm{esc}}$ 
increase from $z\sim 0$ to $6$: 
evolutions of stellar population, outflow, dust extinction, 
and \lya\ scattering of \HI\ in the galaxy's ISM.
It should be noted that the IGM absorption of Ly$\alpha$ becomes
strong from $z\sim 0$ to $6$, and that the evolution of IGM absorption
suppresses $f^{\mathrm{Ly}\alpha}_{\mathrm{esc}}$ (see below for the quantitative arguments),
which cannot be a physical mechanism for the $f^{\mathrm{Ly}\alpha}_{\mathrm{esc}}$ increase
towards high-$z$.
For the possibility of stellar population evolution,
the estimates of the $f^{\mathrm{Ly}\alpha}_{\mathrm{esc}}$ would increase,
if more ionizing photons for a given SFR are produced in galaxies 
that have very massive stars found in the early stage of star-formation.
However, the average/median stellar ages of LAEs for a constant star-formation history are
$10-300$ Myr at $\zi = 2-6$ (c.g., \citealt{Gawiser06, Pirzkal07, Lai08, Ono10a, Ono10b, Guaita11}),
which are comparable with those at $\zi \sim 0$ (e.g., \citealt{Cowie11, Hayes14}).
Because there are no systematic differences in stellar ages by redshift,
the difference of stellar population does not explain the
large increase of $f^{\mathrm{Ly}\alpha}_{\mathrm{esc}}$.
For the possibility of outflow, it is likely that
gas outflow of galaxies help \lya\ photons escape from the
ISM, because the \lya\ resonance wavelength of the ISM
is redshifted by the bulk gas motion of outflow.
If there is a systematic difference in outflow velocities,
the $f^{\mathrm{Ly}\alpha}_{\mathrm{esc}}$ values change.
Because the typical outflow velocities of LAEs 
are $50-200$ km s$^{-1}$ that show no systematic
change over the redshift range
of $z\sim 0-6$ 
\citep{Hashimoto13, Wofford13, Erb14, Shibuya14, Stark15, Rivera15}
\footnote{
These outflow velocity measurements are obtained for UV-continuum bright galaxies,
except for a few lensed galaxies.
Because the outflow velocities of LAEs are similar to those of LBGs (150-200 km s$^{-1}$; 
e.g., \citealt{Hashimoto13, Erb14, Shibuya14}),
UV-continuum faint galaxies would have the outflow velocity comparable to that of UV-bright galaxies.
},
the galaxy outflow would not be a major reason of the large 
$f^{\mathrm{Ly}\alpha}_{\mathrm{esc}}$ increase.
For the possibility of dust extinction evolution,
it is thought that the amount of dust in galaxies decreases from $z \sim 0$ to $6$, 
and that galaxies with small dust extinction have large 
$f^{\mathrm{Ly}\alpha}_{\mathrm{esc}}$ values.
Because the dust attenuation of Ly$\alpha$ is enhanced by
the resonance scattering of \HI\ in the galaxy's ISM that
depends on the \HI\ density, we
first obtain crude estimates of dust extinction effects
with no resonance scattering.
We estimate the luminosity averaged stellar extinction, $E(B-V)_{\star}$, 
from the dust-corrected and uncorrected UV LDs by the equation,
\begin{equation}
\rho^{\mathrm{int, UV}}_{\mathrm{SFRD}} = 10^{0.4 \times E(B-V)_{\star} \times k_{\mathrm{UV}}} \times \rho^{\mathrm{uncorr, UV}}_{\mathrm{SFRD}},
\label{eq:dust}
\end{equation}
where $\rho^{\mathrm{uncorr, UV}}_{\mathrm{SFRD}}$ is
the dust-uncorrected UV SFRDs (Section \ref{sec:evoLD}) calculated with Equation (\ref{eq:sfr_uv}).
The value of $k_{\mathrm{UV}}$ is the extinction coefficient at 1500 \AA,
which is derived with the Calzetti's extinction law \citep{Calzetti00}, $k_{\mathrm{UV}} = 10.3$.
We thus obtain $E(B-V)_{\star}$ values over $\zi \sim 0-6$.
From these $E(B-V)_{\star}$ values, we estimate 
$f^{\mathrm{Ly}\alpha}_{\mathrm{esc, dust}}$ 
with
\begin{equation}
f^{\mathrm{Ly}\alpha}_{\mathrm{esc, dust}} = 10^{-0.4 \times k_{1216} \times E(B-V)_{\rm gas}},	\label{eq:screen1}
\end{equation}
where 
$k_{1216}$ is the extinction coefficient at 1216\AA, $k_{1216} =12.0$, estimated with 
the \citeauthor{Calzetti00}'s law. 
Here we adopt $E(B-V)_{\rm gas} = E(B-V)_\star/0.44$ \citep{Calzetti00}.
The blue open symbols in Figure \ref{fig:LyaLD_evo} present the \lya\ escape fraction values
corrected for dust extinction, $f^{\mathrm{Ly}\alpha}_{\mathrm{esc}}/f^{\mathrm{Ly}\alpha}_{\mathrm{esc, dust}}$,
in the case of no resonance scattering.
The dust-corrected \lya\ escape fractions are nearly unity at $z \sim 6$,
while these fractions significantly drop from $z \sim 4$ to $z \sim 0$. 
At $z \sim 0$, the dust-corrected  \lya\ escape fraction is about an order of magnitude
smaller than unity. There is a clear redshift dependence.
We find that the large fraction of \lya\ escape fraction evolution
can be partly explained by dust extinction with no resonance scattering, but 
that there still remains the large discrepancy at $z<4$.
Thus, the large $f^{\mathrm{Ly}\alpha}_{\mathrm{esc}}$ evolution
requires the evolution of \lya\ scattering of \HI\ in the galaxy's ISM 
from $\zi \sim 0$ to $6$ that enhances the dust attenuation.
Due to the resonance nature of the Ly$\alpha$ line,
an increase of the \HI\ density provides longer path lengths
that strengthen the effects of the ISM scattering 
with a small amount of dust.
Indeed, several studies suggest that the high \HI\ density of
star-forming galaxies largely scatter Ly$\alpha$ photons 
\citep{Shapley03, Pentericci07, Verhamme08, Atek09, Pardy14}.

Here, we estimate the \HI\ column density, $N_{\textsc{Hi}}$, of ISM that needs
to explain the large $f^{\mathrm{Ly}\alpha}_{\mathrm{esc}}$ increase from $\zi \sim 0$ to $6$
with the non-resonant extinction values
obtained by the observational data.
We use the 3D Ly$\alpha$ Monte-Carlo radiative transfer code,
MCLya of \cite{Verhamme06} and \cite{Schaerer11}.
The MCLya code computes the Ly$\alpha$ radiative transfer in an expanding homogeneous shell of
ISM \HI\ and dust that surrounds a central Ly$\alpha$ source.
The dust extinction effects are self-consistently calculated
for the resonance line of Ly$\alpha$.
The MCLya code has
four physical parameters to describe the physical properties of the 
shell: 
$N_{\textsc{Hi}}$,
the nebular dust extinction $E(B-V)_\mathrm{gas}$, 
the radial expansion velocity $v_{\mathrm{exp}}$, and
the Doppler parameter $b$ that includes both 
thermal and turbulent gas motions within the shell.
At each redshift shown in Figure \ref{fig:LyaLD_evo}, we 
derive the best-estimate $N_{\textsc{Hi}}$ value,
using
the $E(B-V)_{\rm gas}$ values obtained above.
We set $b = 12.8$ km s$^{-1}$ that is a fiducial value,
although the $b$ parameter negligibly changes our results.
For $v_{\mathrm{exp}}$, we adopt the average outflow velocity of galaxies at $\zi \sim 0-6$,
$v_\mathrm{exp} = 150$ km s$^{-1}$ \citep{Jones12, Hashimoto13, Shibuya14, Stark15, Rivera15}.
Because the outflow velocity measurements, available to date, have large uncertainties,
we allow the moderately large range of outflow velocities, $v_\mathrm{exp} = 50-200$ km s$^{-1}$,
that includes most of outflow velocity measurements for the low-$z$ and high-$z$ 
LAEs and LBGs so far obtained \citep{Jones12, Hashimoto13, Shibuya14, Stark15, Rivera15}.

\begin{figure}
\centering
\includegraphics[width=8.4cm]{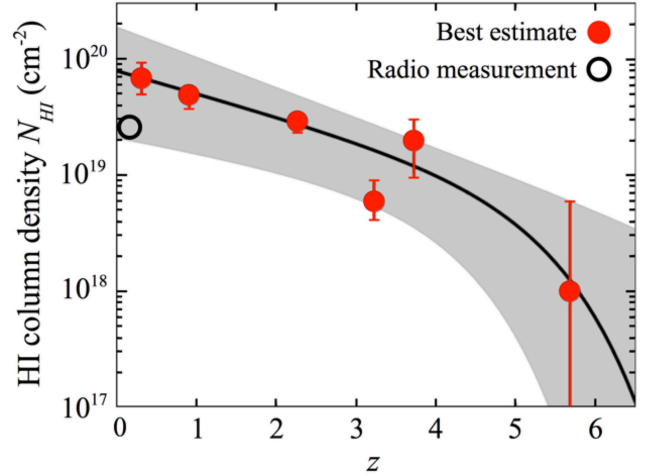}
\caption{Redshift evolution of the \HI\ column density, $N_{\textsc{Hi}}$, of LAEs, as derived
from the 3D Ly$\alpha$ Monte-Carlo radiative transfer code, MCLya.
The red filled circles show the best-estimate $N_{\textsc{Hi}}$ values
for the average outflow velocity of $v_\mathrm{exp} = 150$ km s$^{-1}$ (see text).
The black solid curve is the best-fit function for these $N_{\textsc{Hi}}$ values.
The gray shaded area represents the $N_{\textsc{Hi}}$ range of the best-fit function
allowed for the outflow velocity range of $v_\mathrm{exp} = 50-200$ km s$^{-1}$.
The black open circle denotes the mean $N_{\textsc{Hi}}$ value at $\zi \sim 0.1$
from the radio observations \citep{Pardy14}.
}
\label{fig:NHI_evo}
\end{figure}

We obtain the best-estimate $N_{\textsc{Hi}}$ values
with the three fixed parameters for the MCLya code,
calculating 
the Ly$\alpha$ escape fractions
that agrees with those of the observational estimates. 
For the observational estimates of the Ly$\alpha$ escape fractions,
we use the Ly$\alpha$ escape fraction that is corrected for
the IGM absorption, $f^{\mathrm{Ly}\alpha}_{\mathrm{esc}}/f^{\mathrm{Ly}\alpha}_{\mathrm{esc, IGM}}$
(red open symbols in the bottom panel of Figure \ref{fig:LyaLD_evo}),
where $f^{\mathrm{Ly}\alpha}_{\mathrm{esc, IGM}}$ is 
the Ly$\alpha$ escape fraction contributed only by the IGM \HI\ Ly$\alpha$ absorption.
We estimate $f^{\mathrm{Ly}\alpha}_{\mathrm{esc, IGM}}$ with
the formalism of \citet{Madau95}, assuming no effects of Ly$\alpha$ dumping wing absorption
that is negligible in the redshift range of $z=0-6$ after the cosmic reionization.
In Figure \ref{fig:NHI_evo}, we show the best-estimate $N_{\textsc{Hi}}$ values
for the average outflow velocity of $v_\mathrm{exp} = 150$ km s$^{-1}$.
We fit a function of
$N_{\textsc{Hi}} = n^*(\textit{z}^*/\exp(\textit{z}))^{p} \exp( -\exp(\textit{z})/\textit{z}^*)/\textit{z}^*$
to these $N_{\textsc{Hi}}$ estimates at $z=0-6$,
where 
$p$, $n^*$, and $\textit{z}^*$ are free parameters.
We obtain the best-fit parameters of $n^* = 1.25 \times 10^{21}$ cm$^{-2}$, $p = 0.52$, and $\textit{z}^{*} = 329$.
In Figure \ref{fig:NHI_evo}, the black solid curve represents the best-fit function,
and the gray shaded area exhibits the $N_{\textsc{Hi}}$ range of the best-fit function
that is allowed in the outflow velocity range of $v_\mathrm{exp} = 50-200$ km s$^{-1}$.
Figure \ref{fig:NHI_evo} indicates that $N_{\textsc{Hi}}$ decreases from $\zi \sim 0$ to $6$,
and the best-estimate $N_{\textsc{Hi}}$ values at $\zi \sim 0$, $2$, and $6$ are
$\sim 7 \times 10^{19}$, $\sim 3 \times 10^{19}$, and $\sim 1 \times 10^{18}$ cm$^{-2}$, respectively.
Our $N_{\textsc{Hi}}$ estimates agree with 
the one
obtained by the independent approach of
\cite{Pardy14}, which is presented in Figure \ref{fig:NHI_evo}.
\cite{Pardy14} measure $N_{\textsc{Hi}}$ of star-forming galaxies at 
$\zi \sim 0.1$ 
with the \HI\ imaging and spectroscopic data of the 100m Green Bank Telescope.
\cite{Hashimoto15} 
also estimate $N_{\textsc{Hi}}$ with
\lya\ line profiles of galaxies at 
$\zi \sim 2$ 
based on the high resolution spectra,
and these 
$N_{\textsc{Hi}}$ estimates are 
similar to those of our study.
The agreements 
between our results and 
these studies 
suggest that our 
$N_{\textsc{Hi}}$ estimates are reasonably reliable.
In Figure \ref{fig:NHI_evo},
we find that the $N_{\textsc{Hi}}$ decrease with dust extinction of Ly$\alpha$ resonant scattering
can explain the large $f^{\mathrm{Ly}\alpha}_{\mathrm{esc}}$ increase at $\zi \sim 0 - 6$,
even if we allow the uncertainty of the outflow velocity measurements.
The picture of the $N_{\textsc{Hi}}$ decrease is consistent with 
the increase of the ionization parameter towards high-$z$ suggested by \cite{Nakajima14}.
Because high-$z$ galaxies 
with a high ionization parameter
may 
have density-bounded nebulae
(see Figure 12 of  \citealt{Nakajima14}),
a large fraction of neutral hydrogen in ISM is ionized, which shows a small $N_{\textsc{Hi}}$. 
The $N_{\textsc{Hi}}$ decrease is also consistent with the picture
that the ionizing photon escape fraction increases towards high-$z$
(e.g., \citealt{Inoue06, Ouchi09, Dijkstra14, Nakajima14}).
Our results suggest that the large $f^{\mathrm{Ly}\alpha}_{\mathrm{esc}}$ increase 
is self-consistently explained by the decreasing $N_{\textsc{Hi}}$, which
weakens the ISM dust attenuation through the Ly$\alpha$ resonance scattering.
If we assume the expanding shell models,
the typical $N_{\textsc{Hi}}$ decreases from
$\sim 7 \times 10^{19}$ ($\zi \sim 0$) 
to $\sim 1 \times 10^{18}$ cm$^{-2}$ ($\zi \sim 6$).

%%%%%%%%%%%%%%%%%%%%%%%%%%%%%%%%%%%%%%%%%%%%%%%%%%%%%%%%%%%%%%%%
\section{Summary}\label{sec:summary}

We have conducted the deep and large-area Subaru/Suprime-Cam imaging survey with the narrowband filter, \textit{NB387}.
We have observed five independent blank fields of SXDS, COSMOS, CDFS, HDFN, and SSA22 
whose total survey area is $\simeq 1.43$ deg$^2$.
We make the sample consisting of 3,137 LAEs at $\zi = 2.2$, which is the largest LAE sample, to date,
that is about an order of magnitude larger than the typical LAE samples in previous studies.
The sample covers a very wide \lya\ luminosity range of 
$\log L_{\mathrm{Ly}\alpha} = 41.7-44.4$ erg s$^{-1}$
that allows us to determine bright and faint ends of the \lya\ LFs.
The major findings of our study are summarized below.

\begin{enumerate}

\item
Using our large LAE sample, we derive the \lya\ LFs at $\zi = 2.2$
with small uncertainties including Poisson statistics and cosmic variance errors
(Figure \ref{fig:LyaLF_comp}).
We fit a Schechter function to our best-estimate \lya\ LF at $\zi = 2.2$, and 
obtain the best-fit Schechter parameters of 
$L^{*}_{\mathrm{Ly}\alpha} = 5.29^{+1.67}_{-1.13} \times 10^{42}$ erg s$^{-1}$,
$\phi^{*}_{\mathrm{Ly}\alpha} = 6.32^{+3.08}_{-2.31} \times 10^{-4}$ Mpc$^{-3}$, and 
$\alpha = -1.75^{+0.10}_{-0.09}$
with no priori assumptions in the parameters.
We find that the faint-end slope of the \lya\ LF at $\zi = 2$ 
is steep. The faint-end slope is comparable to that
of UV-continuum LFs at $\zi \sim 2$ 
reported by \cite{Reddy09} and \cite{Alavi14}.

\item
In our best-estimate \lya\ LF at $\zi = 2.2$, we find a bright-end hump at $\log L_{\mathrm{Ly}\alpha} \gtrsim 43.4$ erg s$^{-1}$,
where the \lya\ LF significantly exceeds beyond the best-fit Schechter function (Figure \ref{fig:LyaLF_comp}).
We investigate our LAEs making the bright-end hump
with multiwavelength data of X-ray, UV, and radio 
that are available in the SXDS and COSMOS fields.
We find that all of the LAEs at $\log L_{\mathrm{Ly}\alpha} > 43.4$ erg s$^{-1}$
are detected in the X-ray, UV, or radio band. This result indicates
that this bright-end hump is not originated from the gravitational lensing magnification bias but AGNs.

\item
We identify a moderate but significant increase of the \lya\ LF by a factor of $\lesssim 2$ 
from $\zi \sim 2$ to $3$.
We extend our investigation from $z=2-3$ to $z=0-8$ 
and present the overall evolutionary trends of \lya\ LFs:
the large increase of the \lya\ LFs from $\zi \sim 0$ to $3$,
no evolution of the \lya\ LFs at $\zi \sim 3-6$, and 
the decrease of the \lya\ LFs at $\zi \sim 6$ and beyond.
Calculating the \lya\ LDs by the integrations of these \lya\ LFs,
we show that \lya\ LDs increase nearly by two orders of magnitude
from $\zi \sim 0$ to $3$, and that \lya\ LDs decreases by a factor of $\sim 2$ from $\zi \sim 6$ to $8$
(see also \citealt{Deharveng08, Ouchi08,Konno14}).
This increase at $\zi \sim 0$ to $3$ is significantly faster than the one of UV LDs,
and the decrease at $\zi \gtrsim 6$ is more rapid than the one of UV LDs.

\item
Based on the LAEs with the detection(s) in the X-ray, UV, or radio band,
we derive the AGN UV-continuum LF at $\zi \sim 2$ down to the faint magnitude limit
of $M_{\rm UV}\sim -22.5$. 
We find that our AGN UV LF covers a magnitude range fainter than the previous
studies with an overlap at $M_\mathrm{UV} \simeq -24.8$
with the SDSS DR9 measurements \citep{Ross13} and the 2dF-SDSS results \citep{Croom09},
and confirm that our AGN UV LF agrees well with the SDSS results
at the overlap magnitude.
Fitting the double-power law function to the AGN UV LF data
obtained by our and previous studies,
we constrain the faint-end slope of the AGN UV LF at $\zi \sim 2$, $\alpha_{\rm AGN}=-1.2 \pm 0.1$,
that is flatter than those at $\zi = 4-6.5$, $\alpha_{\rm AGN} \simeq -1.5$ - $-1.8$,
given by \citet{Ikeda11,Giallongo15}.

\item
We estimate $f^{\mathrm{Ly}\alpha}_{\mathrm{esc}}$ values from 
the \lya\ and UV LDs at $\zi \sim 0-8$ given by our and previous studies.
There is a $f^{\mathrm{Ly}\alpha}_{\mathrm{esc}}$ decrease at $\zi \gtrsim 6$ that
can be explained
by the \lya\ scattering of the IGM \HI\ at the EoR.
We find a large $f^{\mathrm{Ly}\alpha}_{\mathrm{esc}}$ increase from $\zi \sim 0$ to $6$
by two orders of magnitude.
This large $f^{\mathrm{Ly}\alpha}_{\mathrm{esc}}$ increase
can be explained neither by stellar population nor outflow
because there exist no significant evolutions in stellar population and 
outflow in LAEs at $\zi \sim 0-6$.
The dust extinction with no Ly$\alpha$ resonance scattering 
can partly explain the $f^{\mathrm{Ly}\alpha}_{\mathrm{esc}}$ increase 
at $\zi \sim 0-6$, but there remains a significantly large discrepancy at $z<4$.
Thus, the Ly$\alpha$ resonance scattering in \HI\ ISM is an important
effect to explain the large $f^{\mathrm{Ly}\alpha}_{\mathrm{esc}}$ increase.
Based on the average $E(B-V)_{\rm gas}$ values for non-resonance nebular lines 
estimated with the observational data,
our simple expanding shell models of MCLya suggest that
the typical \HI\ column density of ISM should decrease from 
$\sim 7 \times 10^{19}$ ($\zi \sim 0$) to $\sim 1 \times 10^{18}$ cm$^{-2}$ ($\zi \sim 6$)
to explain the large $f^{\mathrm{Ly}\alpha}_{\mathrm{esc}}$ increase.

\end{enumerate}

\acknowledgments

We thank Hajime Sugai, Daniel Kunth, Nobunari Kashikawa, Tohru Nagao,
Jorryt Matthee, Shun Saito, and Yuichi Harikane for useful comments and discussions.
This work was supported by World Premier International Research
Center Initiative (WPI Initiative), MEXT, Japan,
and KAKENHI (23244025 and 15H02064) Grant-in-Aid for Scientific Research (A)
through Japan Society for the Promotion of Science (JSPS).
A.K. acknowledges support from the JSPS through the JSPS Research Fellowship for Young Scientists.

\textit{Facility: Subaru} (Suprime-Cam)

%%%%%%%%%%%%%%%%%%%Bibliography%%%%%%%%%%%%%%%%%%%%%%%%%%%%%%%%%%%%%%%
\bibliographystyle{apj}
\bibliography{z2p2LAE_2015}
%%%%%%%%%%%%%%%%%%%%%Table%%%%%%%%%%%%%%%%%%%%%%%%%%%%%%%%%%%%%%%%

\end{document}

%% file: imaging_data.tex
\begin{deluxetable*}{lccccclc}
\tabletypesize{\footnotesize}
\tablecaption{Summary of \textit{NB387} Observations and 
Data
\label{table:imaging_data}}
\tablewidth{0pt}
\tablehead{
\colhead{Band}	 &
\colhead{Field} & 
\colhead{Exposure Time} &
\colhead{PSF FWHM\tablenotemark{$\dagger$}} &
\colhead{Area\tablenotemark{a}} &
\colhead{$m_\mathrm{lim}$\tablenotemark{b}} &
\colhead{Date of Observations}	&
\colhead{Reference\tablenotemark{c}}\\
\colhead{} & 
\colhead{} & 
\colhead{(hr)} &
\colhead{(arcsec)} &
\colhead{(arcmin$^2$)} & 
\colhead{(mag)} &
\colhead{}  &
\colhead{}  
}
\startdata
\textit{NB387}	& SXDS-C				& 3.2		& 0.88	& 587		& 25.7				& 2009 Dec 14$-$16	& (1), (2)	\\
			& SXDS-N				& 2.5		& 0.70	& 409		& 25.6				& 2009 Dec 16			& (1), (2)	\\
			& SXDS-S					& 2.5		& 0.85	& 775		& 25.7				& 2009 Dec 16			& (1), (2)	\\
			& SXDS-E\tablenotemark{d}	& 3.3		& 1.95	& $\cdots$	& $\cdots$			& 2009 Dec 19, 20		& (1), (2)	\\
			& SXDS-W				& 1.8		& 1.23	& 232		& 25.1\tablenotemark{e}	& 2009 Dec 16, 19		& (1), (2)	\\
			& COSMOS				& 4.5		& 0.97	& 845		& 26.1				& 2009 Dec 14$-$16	& (2)		\\
			& CDFS					& 8.0		& 0.85	& 577		& 26.4				& 2009 Dec 14$-$15	& (2), (3)	\\
			& HDFN					& 9.3		& 0.90	& 913		& 26.5				& 2009 Dec 14$-$16	& (2)		\\
			& SSA22					& 1.0		& 0.91	& 800		& 24.9				& 2009 Jul 20			& (2)		\\
\addlinespace[2pt]
\cline{2-8}
\addlinespace[2pt]
			& Total					& 36.1	& $\cdots$& 5138		& $\cdots$			& $\cdots$			& $\cdots$\\
\addlinespace[2pt]
\hline
\addlinespace[2pt]
\multicolumn{8}{c}{Archival Broadband Data}	\\
\addlinespace[2pt]
\hline
\addlinespace[2pt]
$\mathcal{U}$			& SXDS-C		&		& 0.85	&			& 26.9				&					& (4)		\\
					& SXDS-N		&		& 0.85	&			& 26.9				&					& (4)		\\
					& SXDS-S			&		& 0.85	&			& 26.9				&					& (4)		\\
					& SXDS-E			&		& 0.85	&			& 26.9				&					& (4)		\\
					& SXDS-W		&		& 0.85	&			& 26.9				&					& (4)		\\
					& COSMOS		&		& 0.90	&			& 27.2				&					& (5)		\\
					& CDFS			&		& 0.80	&			& 28.0				&					& (6)		\\
					& HDFN			&		& 1.29	&			& 26.4\tablenotemark{e}	&					& (7)		\\
					& SSA22			&		& 1.00	&			& 26.3				&					& (8)		\\
\addlinespace[2pt]
\hline
\addlinespace[2pt]
\textit{B}				& SXDS-C		&		& 0.80	&			& 27.5				&					& (9)		\\
					& SXDS-N		&		& 0.84	&			& 27.8				&					& (9)		\\
					& SXDS-S			&		& 0.82	&			& 27.8				&					& (9)		\\
					& SXDS-E			&		& 0.82	&			& 27.5				&					& (9)		\\
					& SXDS-W		&		& 0.78	&			& 27.7				&					& (9)		\\
					& COSMOS		&		& 0.95	&			& 27.5				&					& (10)		\\
					& CDFS			&		& 0.97	&			& 26.9				&					& (11)	\\
					& HDFN			&		& 0.77	&			& 26.3\tablenotemark{e}	&					& (7)		\\
					& SSA22			&		& 1.02	&			& 26.7				&					& (8)		
\enddata
\tablenotetext{a}{The effective area for the $\zi = 2.2$ LAE selection. 
The effective areas of SXDS-C, -N, -S, -E, and -W are limited by the \textit{u$^*$} image which covers $77$\% of 
SXDS
(see \citealt{Nakajima12} for details).
The area 
of
CDFS is constrained by the deep $U$-band image taken 
with VLT/VIMOS
\citep{Nonino09}.}
\tablenotetext{b}{The $5\sigma$ limiting magnitude in a circular aperture with a diameter of $2\farcs0$.
}
\tablenotetext{c}{(1) \cite{Nakajima12}; (2) \cite{Nakajima13}; (3) \cite{Kusakabe15}; (4) S. Foucaud et al., in preparation (see also \citealt{Nakajima12});
(5) \cite{McCracken10}; (6) \cite{Nonino09}; (7) \cite{Capak04}; (8) \cite{Hayashino04}; (9) \cite{Furusawa08};
(10) \cite{Capak07}; (11) \cite{Hildebrandt06}.}
\tablenotetext{d}{
We do not use the \textit{NB387} image of SXDS-E since the PSF FWHM is relatively large. 
}
\tablenotetext{e}{
We use $2\farcs5$ and $3\farcs0$ diameter apertures for \textit{NB387} of SXDS-W and \textit{UB} of HDFN, respectively, due to bad seeings.
}
\tablenotetext{$\dagger$}{
We homogenize the PSF sizes of broadband and narrowband images in each field (see Section \ref{sec:reduc}).
}
\end{deluxetable*}

%% file: sample.tex
\begin{deluxetable*}{lccccc}
\tabletypesize{\footnotesize}
\tablecaption{Photometric Sample of $\zi = 2.2$ LAEs \label{table:sample}}
\tablewidth{0pt}
\tablehead{
\colhead{Field}	 &
\colhead{All LAE sample\tablenotemark{a}} & 
\colhead{X-ray detection\tablenotemark{b}} &
\colhead{UV detection\tablenotemark{c}} &
\colhead{Radio detection\tablenotemark{d}} &
\colhead{Culled sample\tablenotemark{e}}
}
\startdata
\multicolumn{6}{c}{The full sample}	\\
\addlinespace[2pt]
\hline
\addlinespace[2pt]
SXDS-C				& 277	& 3 [3]		& 3 [3]		& 0 [0]		& 274					\\
SXDS-N				& 239	& 4 [4]		& 5 [4]		& 0 [0]		& 234					\\
SXDS-S				& 374	& 5 [3]		& 5 [4]		& 1 [1]		& 367					\\
SXDS-W\tablenotemark{f}	& 44		& 0 [0]		& 0 [0]		& 0 [0]		& 44					\\
COSMOS				& 642	& 20 [10]		& 10 [10]		& 7 [5]		& 619					\\
CDFS				& 423	& 6 [4]		& $\cdots$	& 6 [4]	 	& 415					\\
HDFN				& 967	& 7 [1]		& 11 [1]		& $\cdots$	& 950					\\
SSA22				& 171	& $\cdots$	& 3 [$\cdots$]	& $\cdots$	& 168					\\
\addlinespace[2pt]
\hline
\addlinespace[2pt]
Total\tablenotemark{g}	& 3137 (1576)	& 45			& 37			& 14 			& 3071 (1538)					\\
\addlinespace[2pt]
\hline
\addlinespace[2pt]
\multicolumn{6}{c}{The EWgt60 sample}	\\
\addlinespace[2pt]
\hline
\addlinespace[2pt]
SXDS-C				& 103	& 2 [2]		& 2 [2]		& 0 [0]		& 101					\\
SXDS-N				& 69		& 0 [0]		& 0 [0]		& 0 [0]		& 69					\\
SXDS-S				& 129	& 1 [0]		& 1 [0]		& 0 [0]		& 127					\\
SXDS-W\tablenotemark{f}	& 6		& 0 [0]		& 0 [0]		& 0 [0]		& 6					\\
COSMOS				& 194	& 9 [4]		& 4 [4]		& 3 [3]		& 184					\\
CDFS				& 142	& 3 [2]		& $\cdots$	& 2 [2]		& 139					\\
HDFN				& 298	& 2 [0]		& 0 [0]		& $\cdots$	& 296					\\
SSA22				& 44		& $\cdots$	& 1 [$\cdots$]	& $\cdots$	& 43					\\
\addlinespace[2pt]
\hline
\addlinespace[2pt]
Total					& 985	& 17			& 8			& 5			& 965				
\enddata
\tablenotetext{a}{The numbers of $\zi = 2.2$ LAE candidates 
after the color selection and rejection of spurious objects.
}
\tablenotetext{b}{The numbers of 
$\zi = 2.2$ LAE candidates
detected in the X-ray data.
The values in square brackets represent the numbers of objects that are also detected in the UV and/or radio data.}
\tablenotetext{c}{The numbers of 
$\zi = 2.2$ LAE candidates
detected in the UV data taken by \textit{GALEX}.
The values in square brackets show the numbers of objects that are also detected in the X-ray and/or radio data.}
\tablenotetext{d}{The numbers of 
$\zi = 2.2$ LAE candidates
detected in the radio data.
The values in square brackets show the numbers of objects that are also detected in the X-ray and/or UV data.}
\tablenotetext{e}{The numbers of 
$\zi = 2.2$ LAE candidates
with no counterpart detection(s) in multiwavelength data of X-ray, UV, and radio.}
\tablenotetext{f}{
The numbers of LAEs are small in SXDS-W.
This is because the limiting magnitude in SXDS-W is brighter than those in the other fields by $\sim 0.5$ mag,
and the effective area of SXDS-W is smaller than those of the other fields by a factor of $\sim 3$ (Table \ref{table:imaging_data}).
The combination of the bright limiting magnitude and the small area reduces the number of LAEs in SXDS-W.
}
\tablenotetext{g}{The total numbers of $\zi = 2.2$ LAE candidates.
The values in parentheses indicate
the total numbers of LAEs found in the SXDS and COSMOS fields.}
\end{deluxetable*}

%% file: param_total_cull.tex
\begin{deluxetable*}{lccc}
\tabletypesize{\footnotesize}
\tablecaption{Schechter Parameters for Full and Culled Samples \label{table:param_total_cull}}
\tablewidth{0pt}
\tablehead{
\colhead{Sample} & 
\colhead{$\alpha$} &
\colhead{$L^{*}_{\mathrm{Ly}\alpha}$} &
\colhead{$\phi^{*}_{\mathrm{Ly}\alpha}$}	\\
\colhead{} & 
\colhead{} &
\colhead{($10^{42}$ erg s$^{-1}$)} &
\colhead{($10^{-4}$ Mpc$^{-3}$)}
}
\startdata
Full\tablenotemark{a,d}			& $-1.75^{+0.10}_{-0.09}$	& $5.29^{+1.67}_{-1.13}$		& $6.32^{+3.08}_{-2.31}$	\\
SXDS+COSMOS/All\tablenotemark{b}			& $-1.87^{+0.10}_{-0.08}$	& $7.83^{+3.22}_{-2.34}$		& $2.99^{+2.26}_{-1.27}$	\\
SXDS+COSMOS/Culled\tablenotemark{c}		& $-1.72^{+0.12}_{-0.11}$	& $4.28^{+1.47}_{-0.99}$		& $7.33^{+3.89}_{-2.83}$	
\enddata
\tablenotetext{a}{
The full sample, which is constructed from the SXDS, COSMOS, CDFS, HDFN, and SSA22 fields.
}
\tablenotetext{b}{
The sample of
LAEs found in the SXDS and COSMOS fields.}
\tablenotetext{c}{
The sample of
LAEs with no multiwavelength counterpart detection(s) in the SXDS and COSMOS fields.}
\tablenotetext{d}{
In the case that we do not include the data at $\log L_{\mathrm{Ly}\alpha} > 43.4$ erg s$^{-1}$ with the bright-end hump for our fitting,
the best-fit Schechter parameters are $\alpha= -1.72 \pm 0.09$,
$L^{*}_{\mathrm{Ly}\alpha} = 4.80^{+1.21}_{-0.86} \times 10^{42} \ \mathrm{erg} \ \mathrm{s}^{-1}$ and
$\phi^{*}_{\mathrm{Ly}\alpha} = 7.40^{+2.84}_{-2.31} \times 10^{-4} \ \mathrm{Mpc}^{-3}$.
}
\end{deluxetable*}

%% file: previous_studies.tex
\begin{deluxetable*}{lcccc}
\tabletypesize{\footnotesize}
\tablecaption{Schechter Parameters of Previous $\zi \sim 2$ LAE Studies\label{table:previous_studies}}
\tablewidth{0pt}
\tablehead{
\colhead{Study} &
\colhead{$\alpha$} &
\colhead{$L^{*}_{\mathrm{Ly}\alpha}$} &
\colhead{$\phi^{*}_{\mathrm{Ly}\alpha}$}	&
\colhead{$\log L_{\mathrm{Ly}\alpha}$ range}	\\
\colhead{} & 
\colhead{} &
\colhead{($10^{42}$ erg s$^{-1}$)} &
\colhead{($10^{-4}$ Mpc$^{-3}$)}	&
\colhead{}
}
\startdata
This work					& $-1.75^{+0.10}_{-0.09}$	& $5.29^{+1.67}_{-1.13}$		& $6.32^{+3.08}_{-2.31}$		& $41.7-44.4$	\\
\cite{Hayes10}				& $-1.49 \pm 0.27$			& $14.5^{+15.7}_{-7.54}$		& $2.34^{+5.42}_{-1.64}$		& $41.3-42.9$	\\
\cite{Blanc11}				& $-1.7$ (fixed)				& $16.3^{+94.6}_{-10.8}$		& $1.0^{+5.4}_{-0.9}$		& $42.6-43.6$	\\
\cite{Cassata11}			& $-1.6 \pm 0.12$			& $5.0$ (fixed)				& $7.1^{+2.4}_{-1.8}$		& $41.2-43.1$ \\
\cite{Ciardullo12}			& $-1.65$ (fixed)			& $2.14^{+0.68}_{-0.52}$		& $13.8^{+1.7}_{-1.5}$		& $42.1-42.7$	\\
\cite{Ciardullo14}			& $-1.6$ (fixed)				& $39.8^{+98.2}_{-16.4}$		& $0.36$\tablenotemark{a}	& $41.9-43.7$	
\enddata
\tablenotetext{a}{
\cite{Ciardullo14} do not show the errors of $\phi^{*}_{\mathrm{Ly}\alpha}$, although
they present the uncertainties of the total number densities of LAEs integrated down to $\log L_{\mathrm{Ly}\alpha} = 41.5$ erg s$^{-1}$,
$\phi_{\mathrm{tot}} = 9.77^{+3.11}_{-2.36} \times 10^{-4}$ Mpc$^{-3}$.}
\end{deluxetable*}

%% file: schechter_evo.tex
\begin{deluxetable*}{lccccc}
\tabletypesize{\footnotesize}
\tablecaption{Best-fit Schechter Parameters and \lya\ Luminosity Densities\label{table:schechter_evo}}
\tablewidth{0pt}
\tablehead{
\colhead{Redshift}	 &
\colhead{$L^{*}_{\mathrm{Ly}\alpha}$} & 
\colhead{$\phi^{*}_{\mathrm{Ly}\alpha}$} &
\colhead{$\rho^{{\mathrm{Ly}\alpha}_{\rm obs}}$\tablenotemark{a}} &
\colhead{Reference}	\\
\colhead{}	 &
\colhead{($10^{42}$ erg s$^{-1}$)} & 
\colhead{($10^{-4}$ Mpc$^{-3}$)} &
\colhead{($10^{39}$ erg s$^{-1}$ Mpc$^{-3}$)} &
\colhead{}	
}
\startdata
0.3		& $0.71^{+0.32}_{-0.29}$		& $1.12^{+2.45}_{-0.61}$		& $0.055^{+0.019}_{-0.014}$		& \cite{Cowie10}				\\
0.9		& $9.22^{+15.6}_{-3.80}$		& $0.12^{+0.18}_{-0.09}$		& $0.165^{+0.067}_{-0.050}$		& \cite{Barger12}				\\
2.2		& $5.29^{+1.67}_{-1.13}$		& $6.32^{+3.08}_{-2.31}$		& $5.93^{+0.23}_{-0.22}$			& This work (Best estimate)		\\
2.2		& $4.87^{+0.83}_{-0.68}$		& $3.37^{+0.80}_{-0.66}$		& $2.17^{+0.13}_{-0.13}$			& This work (EWgt60 sample)	\\
3.1		& $8.49^{+1.65}_{-1.46}$		& $3.90^{+1.27}_{-0.90}$		& $4.74^{+0.46}_{-0.42}$			& \cite{Ouchi08}				\\
3.7		& $9.16^{+2.03}_{-1.67}$		& $3.31^{+1.42}_{-0.98}$		& $4.36^{+0.73}_{-0.63}$			& \cite{Ouchi08}				\\
5.7		& $9.09^{+3.67}_{-2.70}$		& $4.44^{+4.04}_{-2.05}$		& $5.81^{+1.87}_{-1.43}$			& \cite{Ouchi08}				\\
6.6		& $6.69^{+2.51}_{-1.62}$		& $4.17^{+2.70}_{-1.72}$		& $3.86^{+0.86}_{-0.70}$			& \cite{Ouchi10}				\\
7.3		& $3.23^{+25.0}_{-1.63}$		& $2.82^{+17.6}_{-2.70}$		& $1.12^{+2.30}_{-0.68}$			& \cite{Konno14}					
\enddata
\tablecomments{
For $z = 2.2$ (Best estimate), the best-fit Schechter parameters are determined with the full sample (Section \ref{sec:LFat2}),  
while 
for the other cases, $L^{*}_{\mathrm{Ly}\alpha}$ and $\phi^{*}_{\mathrm{Ly}\alpha}$ are derived 
with a fixed value of $\alpha = -1.8$, which is consistent with the best-fit value for our \lya\ LF at $\zi = 2.2$. 
Note that EW$_0$ limits for the selection of LAEs at $\zi = 0.3$, $0.9$, $2.2$ (Best estimate), $2.2$ (EWgt60 sample), $3.1$, $3.7$, $5.7$, $6.6$, and $7.3$
are EW$_0$ $=15$, $20$, $\sim 20 - 30$, $60$, $\sim 60$, $\sim 40$, $\sim 30$, $\sim 10$, and $\sim 0$, respectively.
}
\tablenotetext{a}{\lya\ luminosity densities 
obtained by integrating the \lya\ LF
down to $\log L_{\mathrm{Ly}\alpha} = 41.41$ erg s$^{-1}$.}
\end{deluxetable*}